\documentclass[fleqn,usenatbib]{mnras}

\usepackage{newtxtext,newtxmath}
\usepackage[T1]{fontenc}

\DeclareRobustCommand{\VAN}[3]{#2}
\let\VANthebibliography\thebibliography
\def\thebibliography{\DeclareRobustCommand{\VAN}[3]{##3}\VANthebibliography}

\usepackage{graphicx}	% Including figure files
\usepackage{amsmath}	% Advanced maths commands

\usepackage{amssymb}	% Extra maths symbols
\usepackage{pdflscape}
\usepackage{capt-of}
\usepackage{textcase}
\usepackage{afterpage}
\usepackage{caption}
\usepackage{longtable}
\usepackage{adjustbox}
\usepackage{placeins}
\title[Identification of TP candidates]{Identification of emission-line stars in transition phase from pre-main sequence to main sequence}

\author[Suman et al.]{
Suman Bhattacharyya,$^{1}$\thanks{E-mail: suman.bhattacharyya@res.christuniversity.in}
Blesson Mathew,$^{1}$
Gourav Banerjee,$^{1}$
R. Anusha,$^{1}$
KT Paul$^{1}$
and Sreeja S Kartha$^{1}$
\\
$^{1}$Department of Physics and Electronics, CHRIST (Deemed to be University), Hosur Main Road, Bangalore, India\\
}

\date{Accepted XXX. Received YYY; in original form ZZZ}

\pubyear{2015}

\begin{document}
\label{firstpage}
\pagerange{\pageref{firstpage}--\pageref{lastpage}}
\maketitle

% Abstract of the paper
\begin{abstract}
Pre-main sequence (PMS) stars evolve into main sequence (MS) phase over a period of time. Interestingly, we found a scarcity of studies in existing literature that examines and attempts to better understand the stars in PMS to MS transition phase. The purpose of the present study is to detect such rare stars, which we named as ‘Transition Phase' (TP) candidates -- stars evolving from the PMS to the MS phase. We identified 98 TP candidates using photometric analysis of a sample of 2167 classical Be (CBe) and 225 Herbig Ae/Be (HAeBe) stars. This identification is done by analyzing the near- and mid-infrared excess and their location in the optical color-magnitude diagram. The age and mass of 58 of these TP candidates are determined to be between 0.1--5 Myr and 2--10.5 M$_\odot$, respectively. The TP candidates are found to possess rotational velocity and color excess values in between CBe and HAeBe stars, which is reconfirmed by generating a set of synthetic samples using the machine learning approach.
\end{abstract}

% Select between one and six entries from the list of approved keywords.
% Don't make up new ones.

%%%%%%%%%%%%%%%%%%%%%%%%%%%%%%%%%%%%%%%%%%%%%%%%%%

%%%%%%%%%%%%%%%%% BODY OF PAPER %%%%%%%%%%%%%%%%%%

\section{Introduction}
Star formation of massive stars proceed through well studied, definitive stages such as protostar and pre-main sequence (PMS) phase, before reaching the main sequence (MS). An understanding of the transition from PMS to MS phase is vital to address the astrophysical aspects connected to disc dynamics and rotation rate evolution of stars. Interestingly, a scarcity of studies is found in existing literature that examines and attempts to better understand the stars in PMS to MS transition phase. This dearth of study naturally raises a few pertinent questions such as, `Why is such a scarcity in the studies of stars in transition phase from PMS to MS' and `Which stellar parameter is a clear and definitive indicator to identify such candidates'. We aim to address these problems in the current study by identifying a sample of stars that are most likely in transition from Herbig Ae/Be to classical Be phase. Herbig Ae/Be and classical Be are both instances of a rather uncommon class of stars known as `emission-line' stars, which exhibit emission lines of several elements in their spectra.

A Herbig Ae/ Be (HAeBe hereafter) star is an intermediate mass PMS star (between 2 and 8 M$_\odot$) with a gaseous disc surrounded by a dusty outer envelope \citep{waters1998herbig}. \cite{herbig1960spectra} was the first to classify these stars as a distinct category of objects. The currently accepted definition of HAeBe stars was proposed by \cite{1994The}, \cite{waters1998herbig} and \cite{2003Vieira} incorporating the following two criteria: (a) Pre-main sequence stars with spectral types A–F that exhibit emission lines in their spectra and (b) that exhibit a considerable infrared (IR) excess due to a hot or cool circumstellar dust shell, or a combination of both. \cite{strom1972nature} confirmed the PMS nature of HAeBe stars, suggesting that their ages range from 0.1-1 Myr. On the other hand, a classical Be (CBe hereafter) star is a type of massive B-type main sequence star that is surrounded by a geometrically thin, equatorial, gaseous, decretion disc that orbits the star in Keplerian rotation \citep{Meilland2007}. The existence of such a circumstellar, gaseous disc was first suggested by \cite{1931Struve}. These stars also belong to the luminosity classes III-V. A comprehensive review of the disc formation in CBe stars and the role of mechanisms such as rotation, non-radial pulsation in addressing the `Be phenomenon' is outlined in \cite{2013Rivinius} and \cite{2003Porter}.

Spectra of both CBe \cite[e.g.][]{2021Banerjee, 2011Mathew, 1996Hanuschik, 1982Slettebak} and HAeBe stars \cite[e.g.][]{Alecian2013, Hernandez2004herbig, 1994Boehm, 1992hamannpersson} show emission lines of different elements such as hydrogen, iron, oxygen, helium and calcium. So it is always difficult to distinguish these two type of stars from their spectral features which are common to both classes. It may be noted that a few HAeBe stars show forbidden emission lines of [OI] 6300, 6363 \AA~in their spectra. Hence, if forbidden lines are present in the spectrum, it can be used as a criterion to separate HAeBe and CBe stars since CBe stars do not show forbidden emission lines in the spectra. However, one can distinguish them in terms of IR excess which denotes the excess flux over the continuum in the IR regime. CBe stars show comparatively lower IR excess than HAeBe stars due to the absence of dust in their disc. Now, the pertinent question is, `Can IR excess be used to identify Ae/Be stars in transition from PMS to MS'? Interestingly, analysis of the near-IR color-color diagrams of emission-line stars reveal the existence of stars that lie between the location of HAeBe and CBe stars \citep{2008Mathew}. They are likely to exhibit properties of both these classes. The star 51 Oph is a probable example of this class \citep{Jamialahmadi2015}. According to \cite{Jamialahmadi2015}, ‘this star appears to be a peculiar source in an unusual transitional state’. 51 Oph, thus is the flagship candidate for transition phase stars. This motivated us to conduct a study to identify transition candidates similar to 51 Oph.

In the present study, we analyzed a sample of 2167 CBe and 225 HAeBe stars obtained from the literature and identified a sample of 98 stars evolving from PMS to MS phase. We define this separate category of stars as ‘Transition Phase' candidates (TP candidates hereafter). To the best of our knowledge, this is the first study to detect and characterize TP candidates till date. It is worth noting that another class of well-known objects is referred to in the literature as ‘transitional disc' candidates (TD candidates hereafter). They are quite different from the candidates we provide here, in terms of not only spectral type but also evolutionary phase. TD candidates are primarily a subgroup of T Tauri stars that provide insight into the evolution of protoplanetary disc into planetary systems \citep{Espaillat2014}. These TD candidates show a slight excess in the near and mid-IR regions and a large excess in the far-IR domain \citep{Strom1989, Skrutskie1990}.

The paper is organised as follows: Sect. \ref{sec:Data} discusses the data inventory used for this study. The procedure for identifying the 98 TP candidates is explained in Sect. \ref{sec:color_analysis}. Sect. \ref{sec:justify} describes the verification process of the identified TP candidates using the WISE color-color diagram, \textit{v}sin\textit{i} distribution study and oversampling approach from Machine Learning. Sect. \ref{sec:conclusions} summarises the results of our current investigation.

\section{Data Inventory}
\label{sec:Data}
For our analysis, we compiled a sample of 2167 CBe stars from the published papers available in the literature. Majority of these known CBe stars (1991 stars) are taken from \cite{2016Chen} and \cite{2011neiner}. In addition, we also obtained 176 CBe stars from \cite{Fabregat1996}, \cite{Yudin2001}, \cite{Chauville2001}, \cite{Levenhagen2006}, \cite{Silaj2010}, \cite{Draper2014}, \cite{Raddi2015}, \cite{Lin2015}, \cite{Chojnowski2015}, \cite{Gkouvelis2016}, \cite{Zorec2016}, \cite{Yu2018} and \cite{Dimitrov2018}. 

Likewise, we selected 225 HAeBe stars from \cite{vioque2018gaia} and \cite{Arun2019}, where they listed a large sample of HAeBe stars. Additional samples of HAeBe stars were obtained from \cite{hernandez2005herbig}, \cite{Manoj2006}, \cite{Baines2006}, \cite{Carmona2010}, \cite{Alecian2013} and \cite{Fairlamb2015}.

We chose for this study only stars that had previously been designated as CBe or HAeBe stars in the literature at least once. This allowed us to include definitive CBe and HAeBe stars for our analysis, which is also the updated sample till date. We excluded any CBe or HAeBe stars from the Large Magellanic Cloud (LMC) and Small Magellanic Cloud (SMC) since their metallicity environments are distinct and could affect the statistical outputs of our analysis.

\section{Procedure for identifying the Transition Phase candidates}
\label{sec:color_analysis}

In this section, we explain the method for identifying TP candidates by analyzing their location in the near-IR (J-H){$_0$} versus (H-K{$_S$}){$_0$} diagram and in the optical color-magnitude diagram (CMD). In addition, we made use of the available photometric data in the mid-IR regime to estimate the IR excess emission in the identified TP candidates.

\subsection{Estimating the extinction parameter values for our sample of stars}
\label{sec:extinction}
As a first step, we estimated the extinction parameter (i.e. A$_J$, A$_V$, A$_H$, A$_K$, A$_B$, A$_{W1}$, A$_{W2}$ and A$_{W3}$) values for J, H, K$_S$, B, V, W1, W2 and W3 bands for our sample stars. The E(B--V) values for our stars are calculated using the dust extinction maps of \cite{2019Green}. We adopted the corresponding distances of these stars from Gaia DR2 \cite{bailer2018estimating}. For the present study, we considered R$_V$=3.1 for CBe \citep{Rv2001} and R$_V$=5 for HAeBe stars \citep{Arun2019, Manoj2006, Hernandez2004herbig}, where R$_V$ is the ratio of total-to-selective extinction. 

Next, considering our calculated A$_V$ (A$_V$ = R$_V$ $\times$ E(B--V)) values, we used the conversion relations taken from %Mathis et al. (2016)
\cite{1989Cardelli} for estimating the A$_J$, A$_H$, A$_K$ and A$_B$ values for our sample stars. Similarly, conversion relations adopted from \cite{Wang2019} are used to calculate the respective A$_{W1}$, A$_{W2}$ and A$_{W3}$ values for these stars.

\subsection{Analysis using near-IR color-color diagram}
\label{sec:nir_CCDM}
The IR excess found in the continuum flux distribution of CBe stars is usually attributed to thermal bremsstrahlung emission from free electrons in an ionized, hot, dense circumstellar disc \citep{1974Gehrz, 1977Hartmann}. On the contrary, in the case of HAeBe stars, IR excess emission is due to the thermal re-emission from the dust present in their circumstellar disc \citep{1992Hillenbrand}. IR excess found for HAeBe stars is usually higher than CBe stars since emission due to the presence of dust is more intense than thermal bremsstrahlung. \cite{1984Finkenzeller} claimed that the (H--K$_S$) colors for HAeBe stars will be $>$ 0.4 mag and for CBe stars, H-K$_S$ will be $<$ 0.2 mag. Thus, using this criterion we separate the sample of bona fide CBe and HAeBe stars in the near-IR color-color diagram (CCDm).

\begin{figure*}
    \centering
    \includegraphics[width=140mm,height=140mm,keepaspectratio]{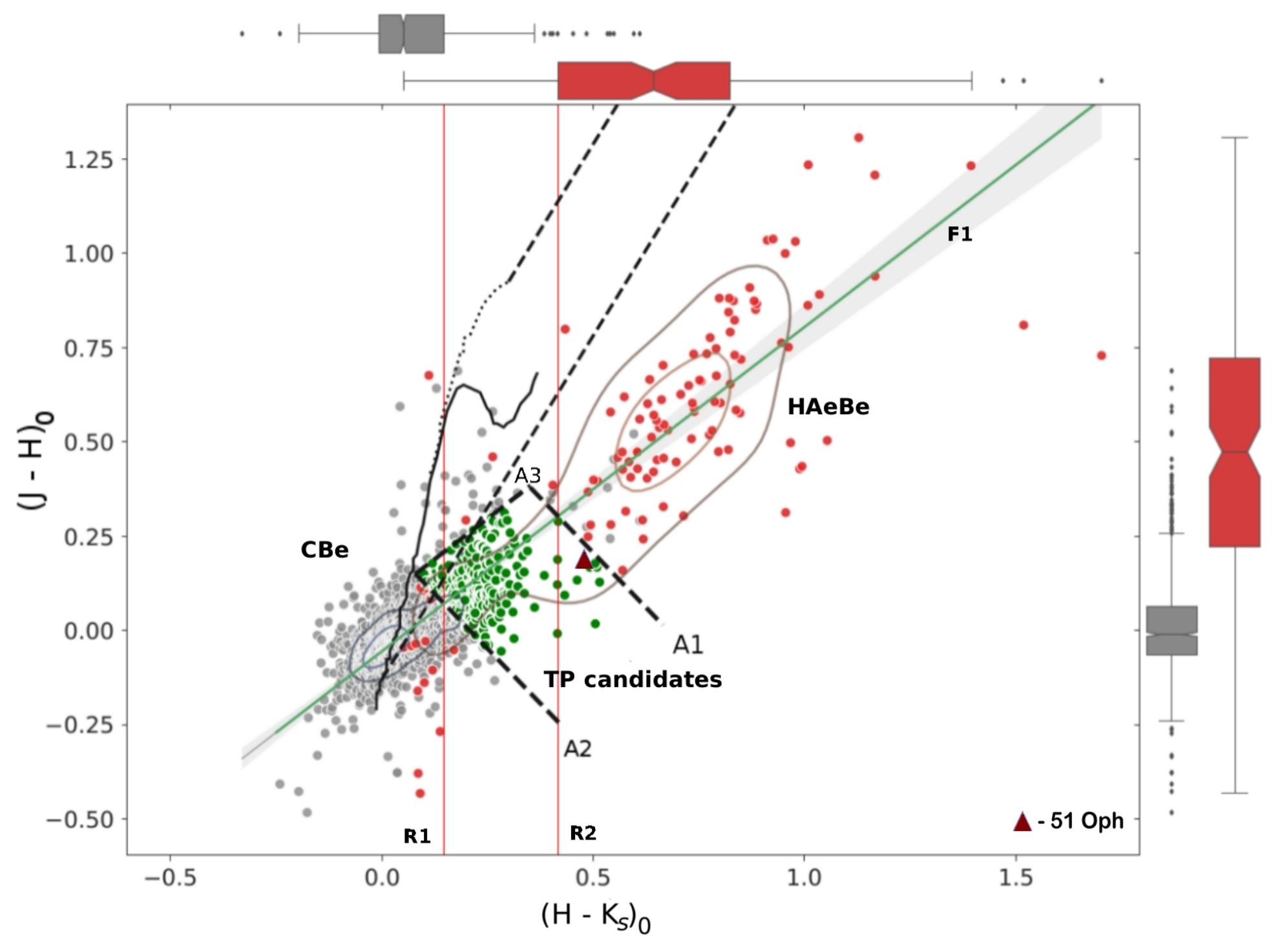}
\caption{Reddening corrected 2MASS color-color diagram of 1319 CBe and 161 HAeBe stars. Here, the grey filled circles represent CBe stars and red filled circles are HAeBe stars. The center of the total distribution of CBe and HAeBe stars is shown by the two vertical, red lines R1 and R2, respectively. Likewise, the green line F1 is the measured linear fit of the total distribution of our sample CBe and HAeBe stars. Green filled circles in the figure represent the estimated Transition Phase (TP) candidates, which are enclosed by the two black, dashed lines A1 and A2. A1 and A2 are drawn perpendicular to F1 for creating a border to our selected probable TP candidates. Another straight line (A3) is drawn parallel to the fitted line F1 and is used to join A1 and A2. This is done to ensure that our selected probable TP candidates do not cross the Main Sequence, which is indicated by the black solid line, and whose data is adopted from \protect\cite{1983Koorneef}. Now, we consider all stars falling in the region encompassed by the three straight lines A1, A2 and A3 as probable TP candidates. The grey contour lines are drawn based on the number density of CBe and HAeBe stars. Thus, from our analysis using the NIR CCDm, we primarily identified a total of 271 probable TP candidates, which are analysed further.}
    \label{fig:ccd1}
\end{figure*}

We queried for the selected 2167 CBe and 225 HAeBe stars in the 2MASS point source catalog \citep{2003Cutri} using VizieR for a search radius of 2$\arcsec$. We found that 2090 CBe and all 225 HAeBe stars have J, H, K$_S$ magnitudes available in the 2MASS catalog. Among them, it was found that 1319 CBe and 161 HAeBe stars have E(B--V) values available in the dust extinction maps of \cite{2019Green}. So we could calculate the A$_V$ values for these stars only. Hence, we considered these stars for further analysis. The relevant correction for extinction of the obtained J, H, K$_S$ magnitudes are done as described in Sect. \ref{sec:extinction}. The reddening corrected 2MASS CCDm of these 1319 CBe and 161 HAeBe stars is shown in Fig. \ref{fig:ccd1}.

It is visible from the figure that our stars are distributed mainly within two regions. The stars marked in red dots are HAeBe stars. We can see that most of them are also satisfying the criterion of H--$K_S$ $>$ 0.4 mag as suggested by \cite{1984Finkenzeller}. Similarly, we also notice that another group of our sample stars are located in a region with H--K$_S$ $<$ 0.2 mag (marked with grey dots), thus agreeing with the criterion for CBe stars as mentioned in \cite{1984Finkenzeller}. Moreover, we notice that a few grey dots are also located in the region of HAeBe stars. Apart from these two observed categories, we also found a number of stars which are falling in the region between H--K$_S$ $>$ 0.2 mag and H--K$_S$ $<$ 0.4 mag. It is expected that the probable TP candidates will lie within this region.

In \cite{1984Finkenzeller}, the sample of HAeBe and ordinary CBe stars were clearly separated in the near-IR color-color diagram (CCDm). Since the sample size is substantially bigger in our study we see an overlapping region for the sample of CBe and HAeBe stars in near-IR CCDm. As a result, it is now important to determine the most likely TP candidates using a proper selection technique.

To begin with we determined the median in the H--K$_S$ axis for both the CBe and HAeBe regions. We found that the median value for the CBe region is 0.05. Similarly, it is 0.64 for the HAeBe region. As a result, we chose only those CBe stars with H--K$_S$ $<$ 0.64 mag and HAeBe stars with H--K$_S$ $>$ 0.05 mag. As an outcome, we acquired 1297 CBe and 140 HAeBe stars for further study.

The midpoint of the total distribution is what is relevant for this study. We chose a region where the H--K$_S$ value for CBe stars is on or above the 75th percentile of the CBe distribution, while the H--K$_S$ value for HAeBe stars is on or below the 25th percentile of the HAeBe distribution. We can see that both directions are pointing towards the middle of the total distribution in this case (depicted by the red vertical lines R1 and R2 in Fig. \ref{fig:ccd1}).

It is imperative that the TP candidates should be within the boundary set by R1 and R2 in Fig. \ref{fig:ccd1}. In order to better filter out the sample, we did a linear fit to the total distribution of CBe and HAeBe stars (shown as the green line F1 in Fig. \ref{fig:ccd1}). This line guided us in selecting a set of stars along the diagonal stretch of the total distribution. We observed the two points where the green line intersected R1 and R2 in Fig. \ref{fig:ccd1}. We formed a border for the TP candidates by projecting two straight lines, A1 and A2, perpendicular to F1 at the intersection points. However, since we are concentrating on candidates that are evolving into MS, our selected candidates should not cross the Main Sequence line (black solid line in the figure, which we adopted from \cite{1983Koorneef}. To do so, we set a constraint by joining A1 and A2 using a third straight line (A3) parallel to the fitted line F1. We considered all stars enclosed within the A1, A2, and A3 boundary lines to be likely TP candidates.

Thus, using the near-IR CCDm, we identified 271 probable TP candidates represented as green dots within the region 0.13 $<$ H–K$_S$ $<$ 0.4 in Fig. \ref{fig:ccd1}.

\subsection{Quantifying IR excess using spectral index}
In the case of young stellar objects (a broader category encompassing PMS and protostars), IR excess is attributed to the presence of circumstellar dust, which absorbs visible light from the young star and re-radiates in the IR region. The closer the dust is to the star, the hotter it is and the shorter is the characteristic wavelength of its infrared emission. If the distribution of dust is away from the star, it radiates at longer wavelengths, in mid- and far-IR \citep{1992Hillenbrand, 1999Malfait}. This emission appears as excess flux over the stellar continuum in the IR region of the Spectral Energy Distribution (SED) plot. The presence of hot/cool dust in the disc of such stars can be determined from the slope of the IR region in the SED.

A classification scheme for young stellar objects (YSOs hereafter) was quantified by \cite{1987Lada} depending on their slope observed in the IR region of the SED, known as Lada indices. Objects are classified into different categories: Class 0, Class I, Class II and Class III depending on the steepness of the indices at various wavelength intervals \citep{1993Andre}. We estimated the n{$_{2-4.6}$} spectral index for 271 probable TP candidates using the 2MASS \textit{K\textsubscript{S}} and WISE \textit{$W2$} magnitudes, which is similar to the index calculated by \cite{Arun2019} for HAeBe stars and \cite{2021anusha} for classical Ae (CAe) stars.

We considered the J, H and K\textit{\textsubscript{S}} magnitudes for analysis of the probable TP candidates as explained in the previous section. To measure the slope we need the data for another second magnitude which lies sufficiently far enough in the wavelength range. For this study we employed K-W2 spectral index, for which we considered the WISE W2 ($\lambda$ = 4.6 $\mu$m) band data obtained from WISE catalog \citep{2014cutri}. We crossmatched our sample of 271 probable TP candidates within a search radius of 2" arcseconds and found that 251 stars have WISE W2 photometric magnitudes available. We consider only these 251 stars for further analysis.

\begin{figure}
    \centering
    \includegraphics[width=\linewidth,height=\textheight,keepaspectratio]{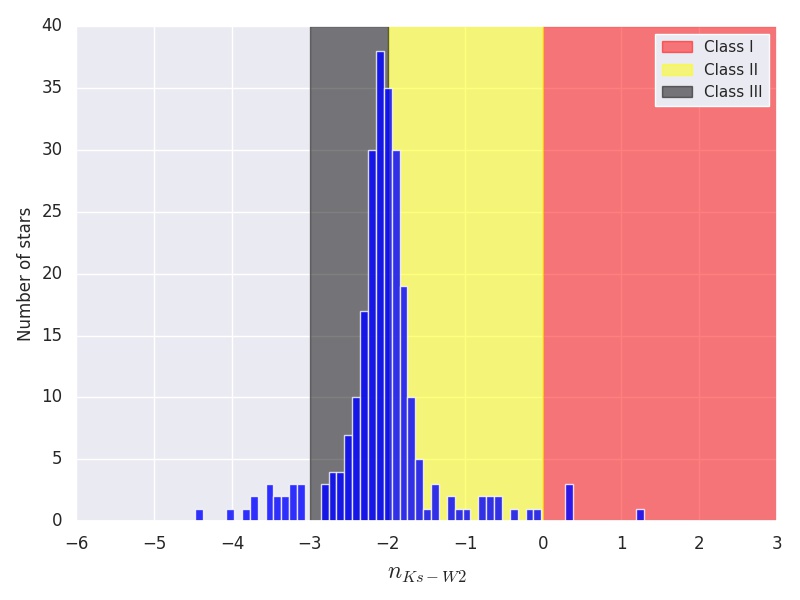}
    \caption{The spectral index distribution for the selected 251 probable TP candidates. Here, the red region represents Class I objects or objects showing high IR excess, the yellow region is for Class II objects which show mediocre IR excess and the grey region represents Class III objects exhibiting either minimum or no IR excess. It is observed  that most of the probable TP candidates are falling on the intersection region of Class II and Class III objects. This confirms that they do not show any distinctive behavior of either CBe or HAeBe stars, but lie in the middle. It is also noticed that the major distribution of the value of spectral indices n{$_{2-4.6}$} varies from -4.0 to 0.4.}
    \label{fig:lada}
\end{figure}

The spectral index distribution of 251 potential TP candidates is depicted in Fig. \ref{fig:lada}. The red region corresponds to Class I objects with a large IR excess, the yellow region corresponds to Class II objects with a lower IR excess than Class I objects, and the grey region corresponds to Class III sources with a low or no IR excess. It is observed from Fig. \ref{fig:lada} that the majority of the candidates fall within the intersection region of Class II and Class III sources. This suggests that the TP candidates show low IR excess, similar to stars losing the dusty disc and evolving to the main sequence. The distribution of the spectral index n{$_{2-4.6}$} is found to be in the range -4.0 to 0.4. We chose the candidates which are above the 25th percentile of the whole distribution for further investigation. We found 187 TP candidates using this selection criterion with spectral index in the range of -2.2 to 1.2. It is found that 8 stars have substantial photometric error in the 2MASS \citep{2003Cutri} and WISE catalogs \citep{2014cutri} and were removed from further analysis. As a result, our sample size reduced to 179 TP candidates.

We did not introduce any cutoff limit on the upper end of the spectral index histogram distribution (Fig. \ref{fig:lada}). This is because 51 Oph belongs to the Class 1 region and needs to be included among our class of objects. However, this star is reported to be a HAeBe star by \cite{Arun2019}, whereas \cite{Jamialahmadi2015} claimed it to show the behaviour of a CBe star. So, by not putting any cutoff on the higher index region, we were able to include such types of peculiar stars also in the sample.

\subsection{Age and Mass estimation of the Transition Phase candidates}
\label{sec:age_mass}
For age and mass measurements, we selected B and V magnitudes of TP candidates from APASS DR9 catalog \citep{2015APASS} to construct the color-magnitude diagram (CMD). We found that 98 among 179 TP candidates have B and V magnitudes and extinction values available. In the present study, we used the extinction corrected M$_V$ versus (B-V)$_0$ CMD to measure the age and mass of these 98 candidates.

We estimated the age and mass of the TP candidates by plotting the Modules for Experiments in Stellar Astrophysics (MESA) isochrones and evolutionary tracks (MIST) \citep{2016choi, 2016dotter} in the APASS CMD. The MIST is an initiative supported by NSF, NASA and Packard Foundation which builds stellar evolutionary models with different ages, masses and metallicities. The updated models in the MIST archive include isochrones and evolutionary tracks for the B and V magnitudes. It is known that HAeBe stars have a range of rotation rates \citep{Alecian2013}. But we adopted the isochrones corresponding to (V/V$_{crit}$ ) = 0.4, since that is the only model available in the MIST database for a rotating system. Also, we adopted the metallicity Fe/H = 0 (corresponding to solar metallicity; $Z_\odot$ = 0.0152) for estimating the age and mass of 98 TP candidates.

\begin{figure*}
    \centering
    \includegraphics[width=\linewidth,height=\textheight,keepaspectratio]{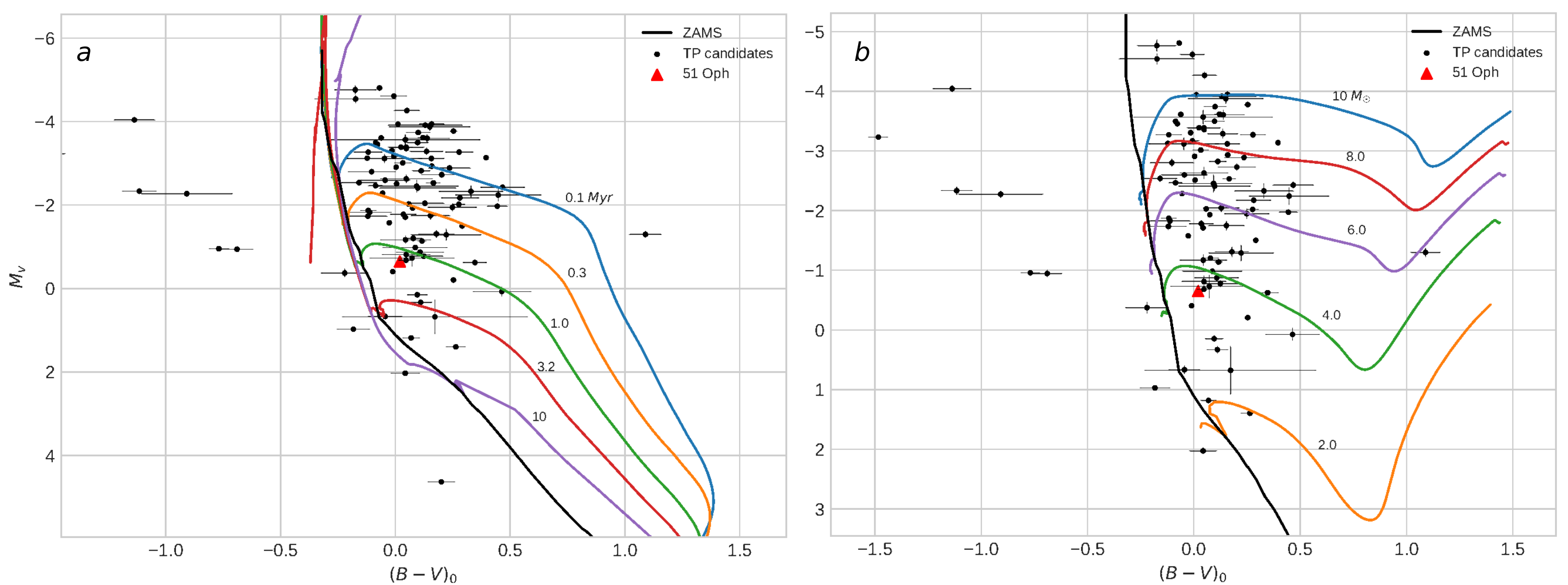}
    \caption{The optical CMD of 98 TP candidates having B and V magnitudes available from APASS \protect\citep{2015APASS}. The black line represents ZAMS from \protect\cite{2013Pecaut}. We estimated the ages and mass of 58 TP candidates by overplotting MIST isochrones and evolutionary tracks. The left panel shows the isochrones, and the right panel shows the evolutionary tracks. The ages and masses are found to be in the range of 0.1 to 5.5 Myr and 2 to 10.5 M$_\odot$, respectively.}
    \label{fig:cmd1}
    \label{fig:cmd2}
\end{figure*}

Fig. \ref{fig:cmd1}a and fig. \ref{fig:cmd2}b show the CMD plot for the 98 most probable TP candidates. From Fig. \ref{fig:cmd1}a, we estimated the ages of 58 TP candidates by overplotting MIST isochrones. Majority of them are found to be in the range of 0.1 to 5.5 Myr. Similarly, from Fig. \ref{fig:cmd2}b, the mass of 58 TP candidates are estimated. Their masses range within 2 to 10.5 M$_\odot$, with the majority of them being greater than 3.5 M$_\odot$. It is also observed that 25 stars are located above the 0.1 Myr isochrone in the CMD. It is possible that they might be in the protostar stage, or are highly evolved stars. We also found 15 other stars located to the left of the main sequence. Photometric variability is seen in HAeBe stars \citep{vandenancher1998} and CBe stars  \citep{2012Paul, 2017bartz}. It is quite possible that the stars to the left of the main sequence may belong to the class of variable PMS stars. Hence, we were unable to measure the age and mass of these 40 (25+15) candidates using the existing MIST isochrones and evolutionary tracks. The list of 98 TP candidates from the present study is presented in Table \ref{tab:table1}.

\section{Verifying the nature of the identified Transition Phase candidates}
\label{sec:justify}

\subsection{Analysis using the WISE color-color diagram}
We used the ALLWISE survey \citep{2014cutri} to investigate the mid-IR properties of the identified 98 TP candidates. The extinction correction for W1, W2 and W3 AllWISE bands are performed as described in Sect. \ref{sec:extinction}.

\begin{figure}
    \centering
    \includegraphics[width=\linewidth,height=\textheight,keepaspectratio]{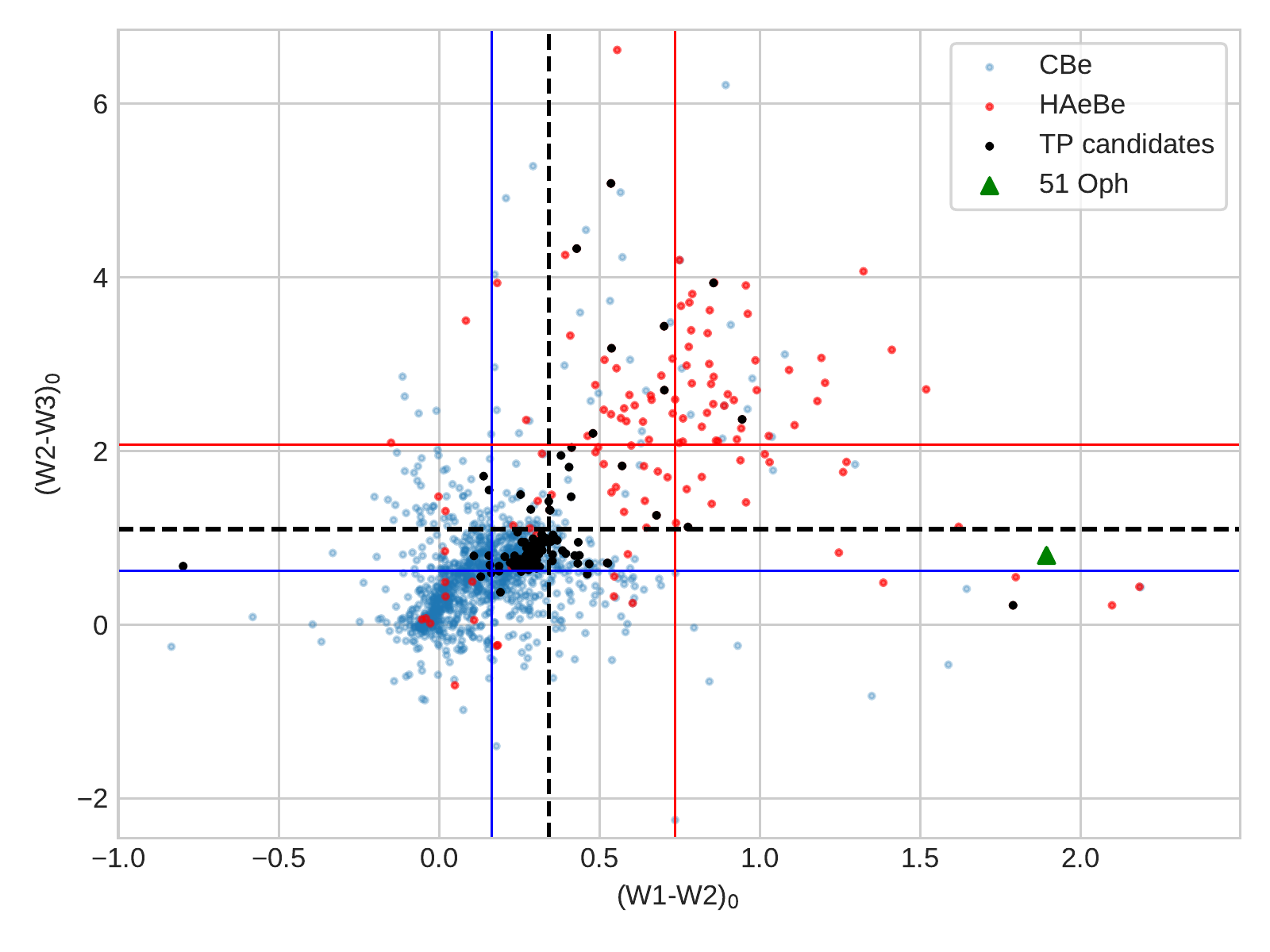}
    \caption{WISE color-color diagram for the 98 TP candidates. Here, black dots represent the TP candidates, whereas red and blue dots represent HAeBe and CBe stars respectively. The mean of the TP candidates, CBe and HAeBe star distributions are depicted by the intersection of the black dashed line, blue and red lines, respectively. The green triangle represents 51 Oph. The comparison of (W1 -- W2)$_0$ and (W2 -- W3)$_0$ colors have been performed with CBe \citep{2016Chen} and HAeBe stars \citep{vioque2018gaia}. We found that most of the TP candidates are located around the upper-right region of CBe stars. This region is also noticed to be overlapping between the regions of the HAeBe and CBe stars.}
    \label{fig:wise}
\end{figure}

The WISE color-color diagram of 98 TP candidates is shown in Fig. \ref{fig:wise}. In the figure, black dots represent the TP candidates, whereas red and blue dots represent HAeBe \citep{vioque2018gaia} and CBe stars \citep{2016Chen}, respectively. We found that most of the TP candidates are located in the upper-right region of CBe stars. The mean of the TP candidates, CBe and HAeBe star distributions are depicted by the intersection of the black dashed line, blue and red lines, respectively. This indicates that most of the TP candidates are showing comparatively higher mid-IR excess than the mean of the CBe distribution. Moreover, 22 TP candidates are situated well within the CBe star region near to (W1 -- W2)$_0$ = 0 and (W2 -- W3)$_0$ = 0, thus showing less or no excess. Apart from these, another one of the identified TP candidates is located at far lower-right region in the figure showing higher excess in (W1 -- W2)$_0$ color but less excess in (W2 -- W3)$_0$ color in comparison with HAeBe stars. This may be due to the emission from enhanced silicate at wavelengths 8-12 $\mu$m and polycyclic aromatic hydrocarbons (PAH) at 3.3 and 11.3 $\mu$m, which greatly influence W1 (3.4 $\mu$m) and W3 (12 $\mu$m) bands, respectively \cite{2016Chen}. We also found 2 another stars which are lying in the region showing (W2 -- W3)$_0$ excess. Spectroscopic investigation is necessary to better understand these stars.

\cite{2016Chen} suggested that the upper-left region in the WISE color-color diagram occurs due to free-free or bound-free emission originating from proton-electron scattering. On the other hand, sources in the lower-right region of the diagram should have circumstellar dust in the envelope. It is visible from Fig. \ref{fig:wise} that the TP candidates are located in the overlapping region of the CBe and HAeBe stars. Through their analysis, \cite{2016chenherbig} classified the CBe region to be within the range of -0.5 $<$ (W1 - W2)$_0$ $<$ 1.5 and -0.5 $<$ (W2 – W3)$_0$ $<$ 3 (Fig. 5 of \cite{2016chenherbig}), and for HAeBe stars it is 0 $<$ (W1 - W2)$_0$ $<$ 2 and 0 $<$ (W2 - W3)$_0$ $<$ 4 (Fig. 6 of \cite{2016chenherbig}). We found that majority of the identified TP candidates are lying within a range of 0.2 $<$ (W1 - W2)$_0$ $<$ 0.5 and 0.6 $<$ (W2 - W3)$_0$ $<$ 1.1, which is well within the overlapping region of CBe and HAeBe stars.

Our analysis demonstrated that the mid-IR excess of TP candidates is less than that of HAeBe stars and more than CBe stars. This is suggestive of the fact that these stars are losing dusty disc and evolving into the main sequence phase, thereby supporting the near-IR analysis and their identification as TP candidates.

\begin{figure*}
    \centering
    \includegraphics[width=120mm,height=150mm,keepaspectratio]{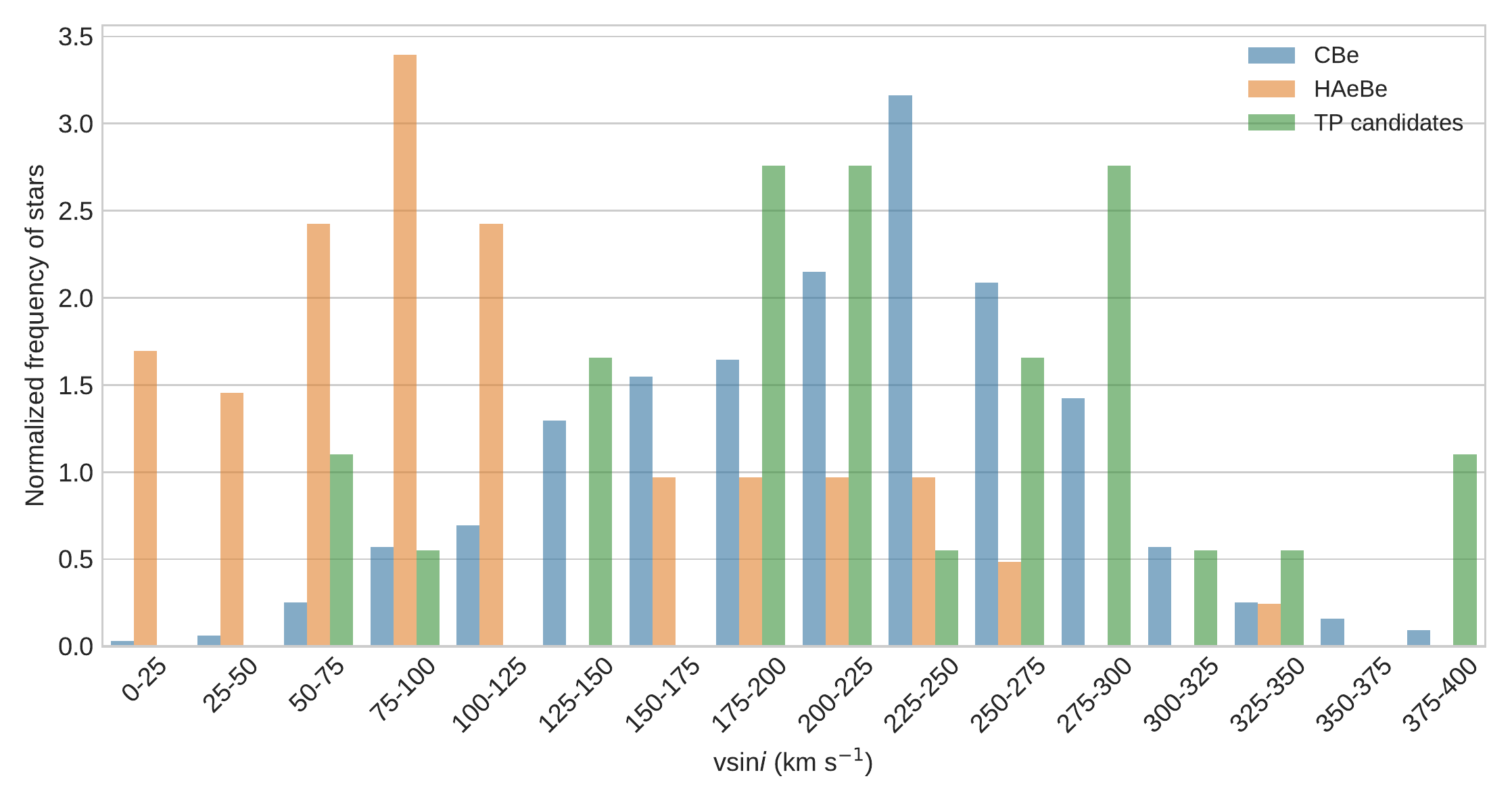}
    \caption{The \textit{v}sin\textit{i} distribution for 460 CBe, 77 HAeBe and 21 TP candidates as obtained from literature. It is clearly visible that HAeBe and CBe stars show distinct peaks, whereas TP candidates are exhibiting multiple peaks. We found that for HAeBe stars \textit{v}sin\textit{i} ranges within 50 – 125 km s\textsuperscript{-1}, whereas in case of CBe stars \textit{v}sin\textit{i} ranges within 175 – 300 km s\textsuperscript{-1}, respectively. For TP candidates \textit{v}sin\textit{i} lies between 175 -- 225 km s\textsuperscript{-1} and 275 -- 300 km s\textsuperscript{-1}. This result shows that \textit{v}sin\textit{i} for the TP candidates are greater than the HAeBe stars and lesser / equal to the upper range of CBe stars.}
    \label{fig:Vsini1}
\end{figure*}

\subsection{Analysis of the \textit{v}sin\textit{i} distribution for our sample}
\label{sec:first_vsini}
To investigate the projected rotational velocity (\textit{v}sin\textit{i}) distribution of TP candidates in comparison with that of CBe and HAeBe stars, we plotted the \textit{v}sin\textit{i} distribution of CBe, HAeBe and TP candidates in Fig. \ref{fig:Vsini1}. The \textit{v}sin\textit{i} values of 460 CBe stars are taken from \cite{Yudin2001} whereas that of 77 HAeBe is obtained from \cite{Alecian2013}. We found \textit{v}sin\textit{i} values of 21 TP stars in the literature and they are provided in Table \ref{tab:table1}.

It is expected that HAeBe stars must lose their dusty discs to evolve into CBe stars. So the system will spin up as a natural consequence of angular momentum conservation. Hence, in a \textit{v}sin\textit{i} distribution plot, we may expect to see a difference between the peak occurrence. The peak should be at lower \textit{v}sin\textit{i} range for HAeBe and at higher \textit{v}sin\textit{i} range for CBe stars. From Fig. \ref{fig:Vsini1}, it is clearly visible that HAeBe and CBe stars actually show distinct peaks, whereas TP candidates are exhibiting bimodal distribution. We found that for HAeBe stars \textit{v}sin\textit{i} ranges within 50 – 125 km s\textsuperscript{-1}, whereas in case of CBe stars \textit{v}sin\textit{i} ranges within 175 – 300 km s\textsuperscript{-1}, respectively. For TP candidates \textit{v}sin\textit{i} show bimodal distribution with peaks at 175 – 225 km s\textsuperscript{-1} and 275 – 300 km s\textsuperscript{-1}, respectively. This result shows that \textit{v}sin\textit{i} for the TP candidates are greater than the HAeBe stars and lesser or equal to the upper range of CBe stars.

As illustrated in Fig. \ref{fig:Vsini1}, the \textit{v}sin\textit{i} distribution of TP candidates is skewed toward CBe stars. This is self-evident, given that the sample of known CBe stars is significantly larger than the sample of known HAeBe stars. As a reason, the selection technique should be influenced by the same skewness towards CBe. In order to mitigate this effect, we conducted an additional analysis employing oversampling using a machine learning technique.

\subsection{Oversampling study using Machine Learning approach}
\label{sec:ML}

Applying Machine Learning techniques to a variety of astronomical problems has gained considerable attention in recent years, particularly when dealing with large datasets for galaxy classification \citep{2001bazellgalaxy}, stellar classification \citep{2002philipstar}, studying the properties of astronomical objects \citep{2004zhangastro} etc. These research make use of a variety of algorithms. Consequently, in the majority of circumstances, the imbalanced class problem tends to bias the learning process in favour of the majority class.

In the present study we found that the \textit{v}sin\textit{i} distribution of TP  candidates is skewed towards large values due to difference in the sample of CBe and HAeBe stars used in the analysis. The number of CBe stars having \textit{v}sin\textit{i} information is about 6 times that of HAeBe stars. Thus, to mitigate the imbalanced character of the datasets and to explain our prior selection criteria, via the near-IR color-color diagram (given in Sect. \ref{sec:nir_CCDM}), we employed SMOTE (Synthetic Minority oversampling Technique) \citep{2002chawla}, a technique that is well-known in the Machine Learning (ML) community. \cite{2020cardio} utilized SMOTE for Cardiovascular Disease (CVD) database, which may enhance the accuracy of CVD detection rate. This technique is also used in astrophysics to de-noise pulsar signals \citep{2018smotepulser}, variable star detection \citep{2020variable}, etc. SMOTE was used in each of these instances due to the unbalanced nature of their dataset, which prompted us to apply SMOTE to our analysis.

The basic strategy is to produce a homogeneous dataset by generating artificial vectors from the minority sample feature space (HAeBe stars in this case). To implement SMOTE, the k-Nearest Neighbour (KNN) supervised machine learning algorithm is utilised to generate synthetic features \citep{knnGou}.

k-Nearest Neighbours are determined for every feature vector in the minority sample, following which, random feature vectors are interpolated between those determined neighbours. Consider `x' as a random sub-sample of the overall minority sample `A'. To determine the k-nearest neighbours (eqn. \ref{eqn:eqn1}), the euclidean distance between `x' and all other candidates in sample A is measured. Then, using a Python random number generator, random samples between 0 and 1 are selected. This is denoted as  \textit{rand}(0,1) in the equation. The random numbers are then multiplied by the distance between the initial vector x and a randomly chosen candidate from the sample within x's neighbour (x$_k$). The information about the shape/distribution of the sample is preserved and transferred over to the newly generated synthetic sample using this procedure.

\begin{equation}
\label{eqn:eqn1}
    {x'}={x+{rand(0,1)*{|x-{x_k}|}}}
\end{equation}

Further, we used the k-fold cross-validation (CV) method to test the newly created sample. At first, the entire distribution is split into k partitions. Then, using (k-1) folds for training, this procedure is done k times. Each iteration includes a test of the training model on the remaining fold. This process is repeated in order to obtain a score referred to as the `accuracy score'. Our needed output is calculated as the average of these scores.

We selected the \textit{v}sin\textit{i} values and 2MASS J, H, K\textit{\textsubscript{S}} magnitudes of 352 CBe stars from \cite{Yudin2001}, 64 HAeBe stars from \cite{Alecian2013} and 21 TP candidates for this oversampling study. Additionally, because CBe and HAeBe datasets are separate from the prior one (in Sect. \ref{sec:nir_CCDM}), we can investigate the statistics independent of our first analysis, using a near-IR color-color diagram (in Sect. \ref{sec:nir_CCDM}) and \textit{v}sin\textit{i} distribution (in Sect. \ref{sec:first_vsini}). The selected feature vectors are J-H, H-K$_S$ and \textit{v}sin\textit{i}. This analysis is performed in three stages. In the final stage we compared our new synthetic output with previous results obtained from Sect. \ref{sec:nir_CCDM} and Sect. \ref{sec:first_vsini}. The three stages are explained as follows:

(i) Initially, we oversampled the HAeBe sample in relation to CBe in order to achieve a close match in the sample ratio. The k-Fold CV method is used to get the best fit and to determine the optimal ratio for maximum accuracy.

(ii) Similar to the first step, we oversampled the TP candidates (21 stars) in comparison to CBe and the new HAeBe list of stars. Once again, the k-Fold CV approach is employed to determine the best ratio between the three samples (CBe, HAeBe and TP candidates).

(iii) From the new list we observed the statistical features and compared them with our analysis performed in Sects. \ref{sec:nir_CCDM} and \ref{sec:first_vsini}.

We obtained a maximum accuracy score of 95\% by performing a 20-fold CV during oversampling of HAeBe stars using a 1:2 ratio of HAeBe to CBe stars. This resulted in a sample size of 176 HAeBe stars, compared to the real sample size of 64. Now, using the CBe and new HAeBe samples, we oversampled our 21 TP candidates. The total number of TP candidates obtained in the new list is 352 (equivalent to CBe), with an accuracy of up to 80\% when employing a 30-fold CV. Additionally, the accuracy score for TP candidates saturates at a maximum of 80\% when the oversampling ratio between TP candidates and CBe stars is 1:2 or above. This score has been maintained at a ratio of up to 1:1 for TP candidates. As a result, we chose the highest number of synthetic TP candidates possible for our study.

\begin{figure}
    \centering
    \includegraphics[width=\linewidth,height=\textheight,keepaspectratio]{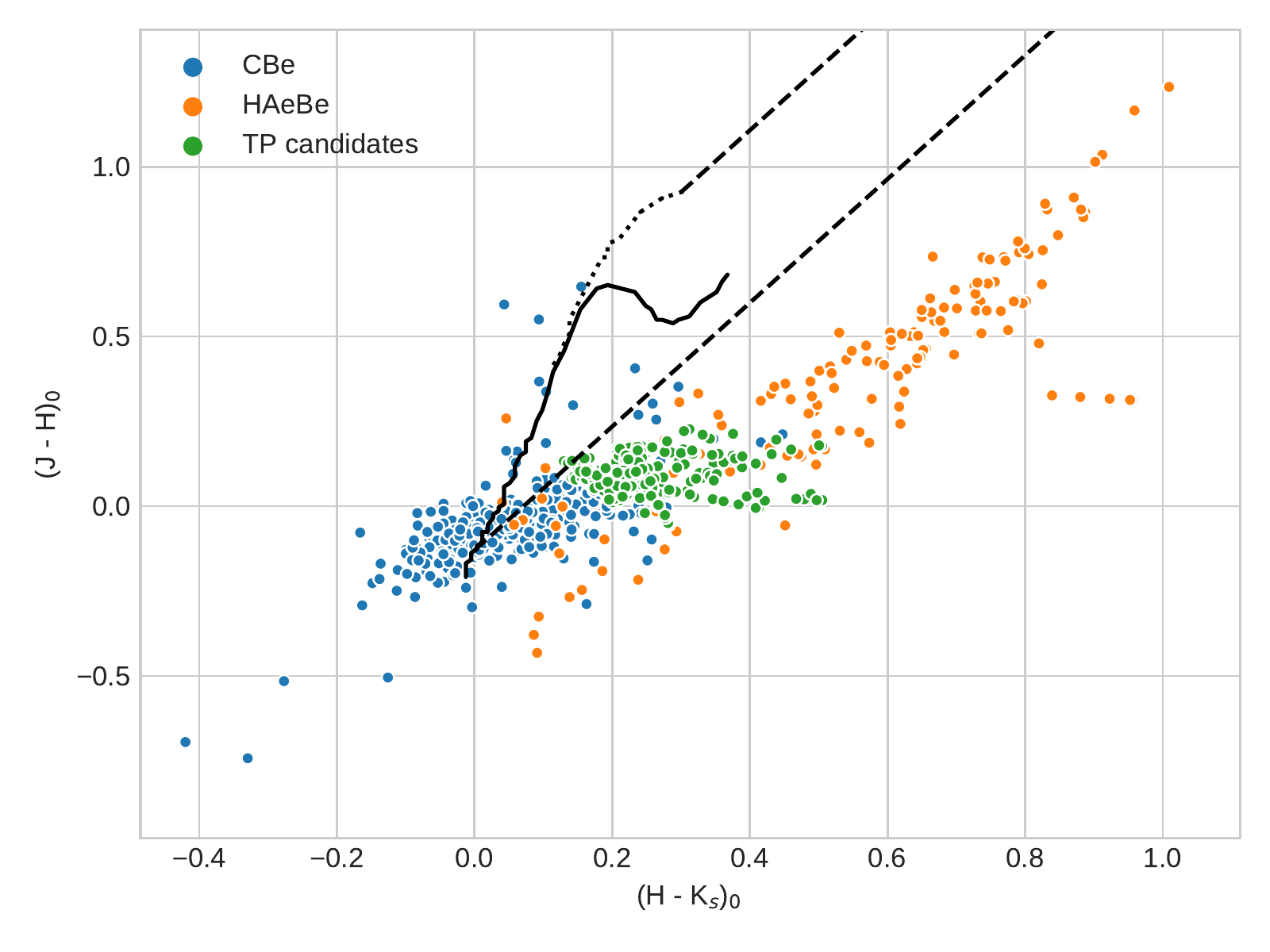}
\caption{The near-IR color-color diagram of 352 CBe and synthetically oversampled 176 HAeBe and TP candidates obtained using the new oversampled dataset. It is clearly evident from the plot that the TP candidates are well positioned between the distribution of CBe and HAeBe stars. It is observed that most of the TP candidates are falling in the region between 0.12 $<$ H-Ks $<$ 0.41, which is 93 percent accurate with respect to our previous estimation for TP candidates. This result satisfactorily verifies our previous statistical selection criteria and shows that the range of TP candidates did not change much from what we found from the previous CCDm (Fig. \ref{fig:ccd1}). The main-sequence, giant branch, and the reddening vectors are represented by solid, dotted, and dashed lines, respectively, as adopted from \protect\cite{1983Koorneef}.}
    \label{fig:smote_CCDm}
\end{figure}

We generated the near-IR color-color diagram and \textit{v}sin\textit{i} distribution plots using the new dataset and compared them with our previous TP candidate estimation. The near-IR color-color diagram obtained with the new oversampled dataset is shown in Fig. \ref{fig:smote_CCDm}. We observed that the majority of TP candidates fall within the range of 0.12 $<$ H-Ks $<$ 0.41, which is 93 percent accurate in comparison to our previous TP candidate estimation. This result validates our prior statistical selection criteria satisfactorily. However, it may be noted that a more balanced, real dataset is required to improve the accuracy of machine learning cross validation.

\begin{figure}
    \centering
    \includegraphics[width=85mm,height=62mm]{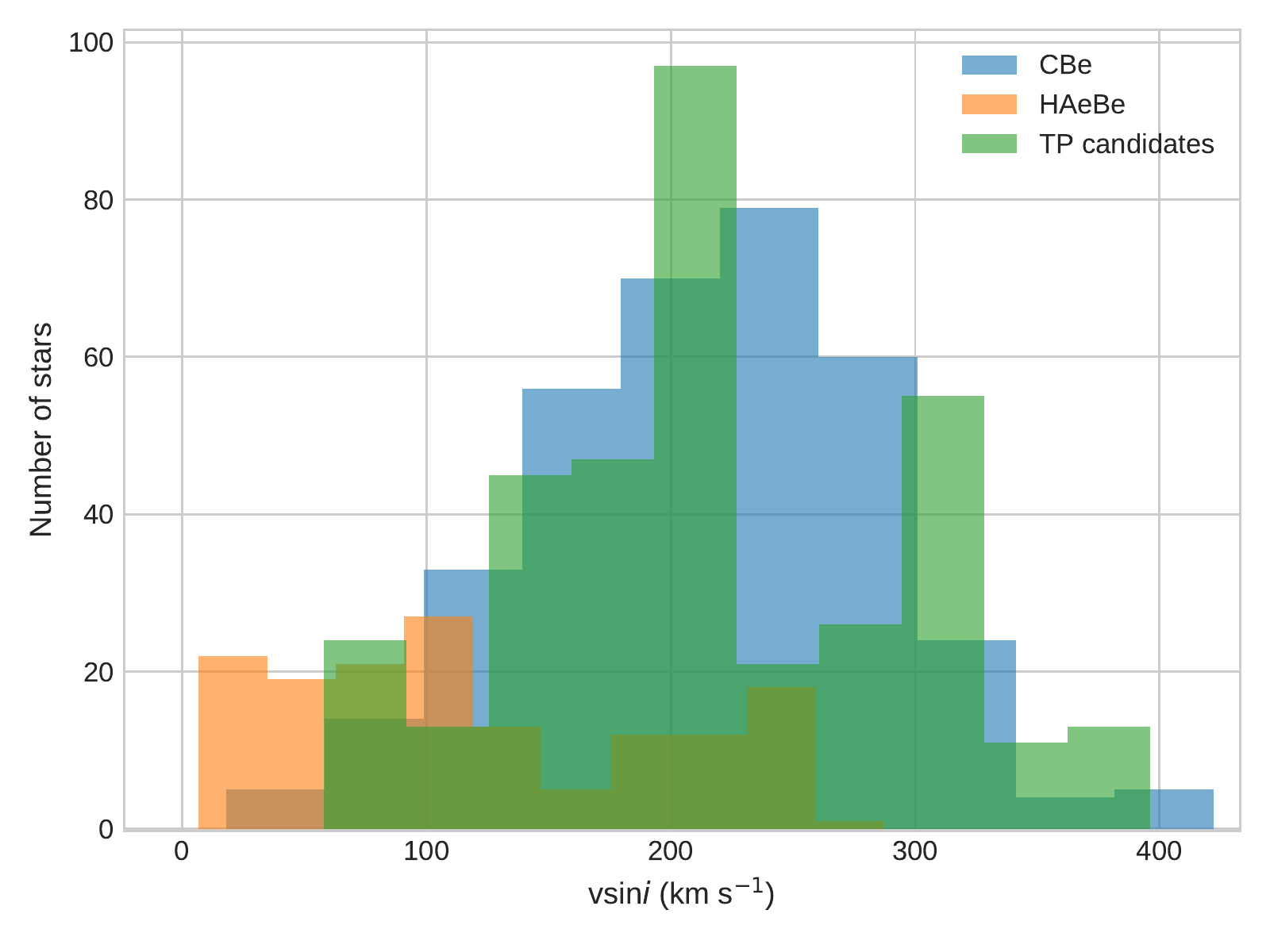}
    \caption{The \textit{v}sin\textit{i} distribution for 352 CBe, 176 HAeBe and 352 TP candidates as obtained using the new oversampled dataset is shown. Similar to what we noticed in Fig. \ref{fig:Vsini1}, the \textit{v}sin\textit{i} peaks of HAeBe and CBe are distinct and are situated in different locations here also. But unlike Fig. \ref{fig:Vsini1}, TP candidates are now exhibiting a single major peak. While for HAeBe and CBe stars, \textit{v}sin\textit{i} range within 50 -- 125 km s\textsuperscript{-1} and 175 -- 300 km s\textsuperscript{-1}, for the new synthetically generated TP candidates \textit{v}sin\textit{i} lies majorly around 200 km s\textsuperscript{-1}. This result shows that \textit{v}sin\textit{i} for the TP candidates have shifted slightly towards the intermediate region of HAeBe and CBe stars, when compared to Fig. \ref{fig:Vsini1}.} 
    \label{fig:Vsini2}
\end{figure}

Fig. \ref{fig:Vsini2} plots the histogram of the \textit{v}sin\textit{i} distributions for 352 CBe, 176 HAeBe, and 352 TP candidates collected from the newly oversampled dataset. The histogram in this figure is very similar to that in Fig. \ref{fig:Vsini1}. As with Fig. \ref{fig:Vsini1}, the \textit{v}sin\textit{i} peaks of HAeBe and CBe are located in distinct positions in Fig. \ref{fig:Vsini2}. It is interesting to see that the \textit{v}sin\textit{i} distribution of TP candidates peak close to the CBe distribution. Unlike in Fig. \ref{fig:Vsini1}, we observed that the major peak is now located around 200 km s\textsuperscript{-1} for TP candidates. This indicates that the peak in Fig. \ref{fig:Vsini2} has shifted slightly towards the lower \textit{v}sin\textit{i} range of the CBe distribution, or toward the HAeBe region, in comparison to Fig. \ref{fig:Vsini1}. The synthetically generated sample exhibits the major distribution characteristics of HAeBe stars and TP candidates. In this context, the larger is the minority sample, more accurate will be the over sampling outcome.

It is quite possible that TP candidates are fast rotators as evident from the analysis. This goes naturally with our expectation that HAeBe stars spin up before getting to the main sequence and hence TP phase of these stars should be fast rotators. However, rotation velocity estimation from the high resolution spectrum of TP candidates and further analysis can only provide confirmation to this hypothesis.

\subsection{Comparison of the rotation velocity distribution of TP candidates with B-type stars}
We assumed in the preceding section that a HAeBe star will pass on to the CBe phase and the identified TP candidates occur somewhere in between such transition. \cite{2010Huang} concluded that late B-type stars (4 M$_\odot$) require a higher V$_{eq}$/V$_{crit}$ threshold to become Be stars than early B-type stars. In other words, higher mass B-type stars have a greater likelihood of becoming CBe stars than low mass B-type stars. Additionally, it is identified that about 10 - 20 \% of B-type stars in the Galaxy go through the classical Be phase during their lifetime \citep{1983Jaschek,1997Zorec,2008Mathew,2017Arcos}. Thus, statistically, there is a possibility that HAeBe stars could evolve to B-type stars rather than CBe stars. Since TP candidates are defined to be in the transition phase from HAeBe to CBe, it is expected that they could evolve to B-type stars rather than CBe stars. Hence, we compared the \textit{v}sin\textit{i} distribution of our newly identified TP candidates with that of a sample of B-type stars and CBe stars.

\begin{figure}
    \centering
    \includegraphics[width=\linewidth,height=\textheight,keepaspectratio]{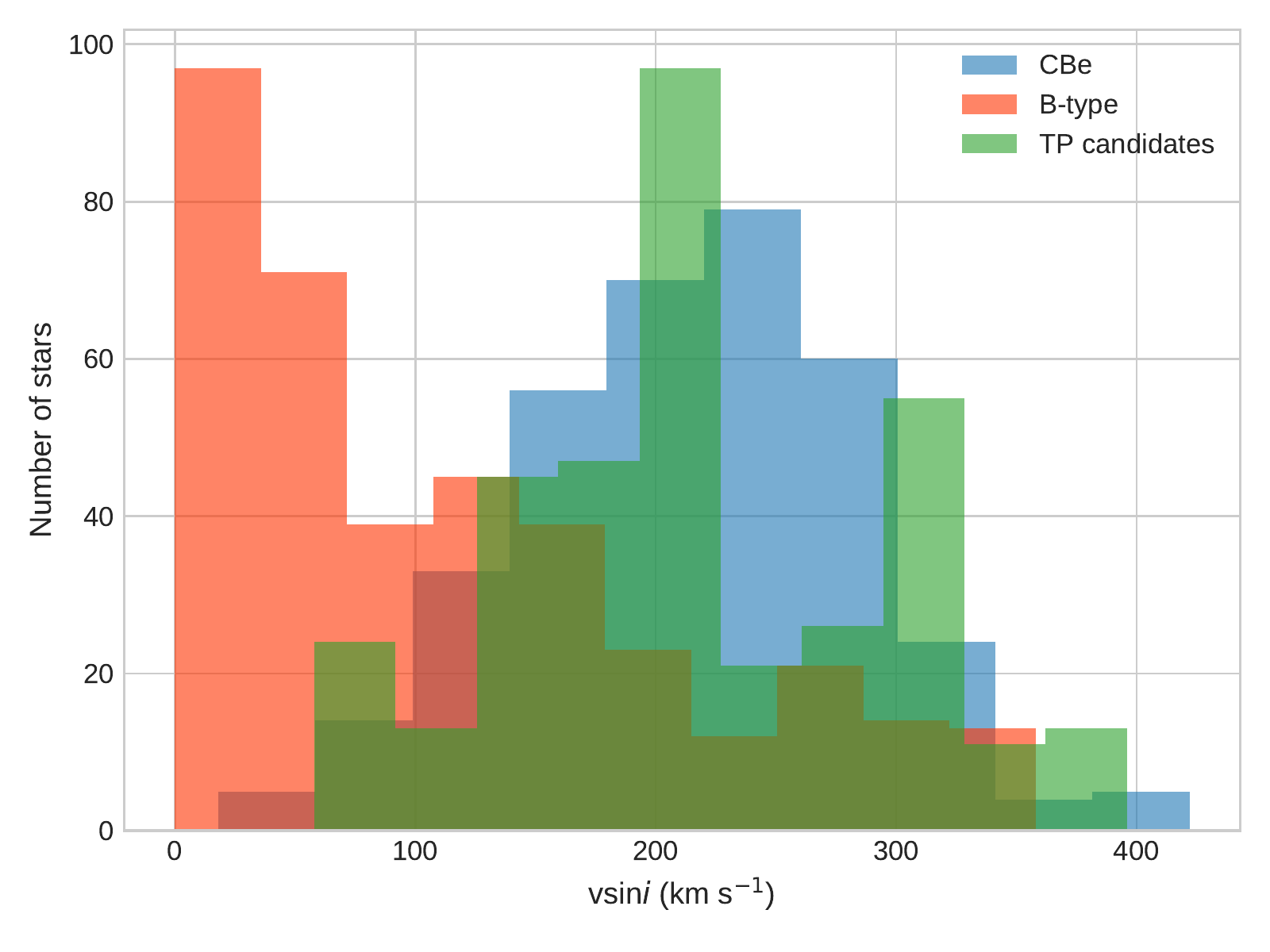}
    \caption{The \textit{v}sin\textit{i} distribution of 376 B-type, 352 CBe stars and 352 synthetic TP candidates are shown in the figure. We observed that B-type stars with high \textit{v}sin\textit{i} values coincide with the distribution of CBe stars and TP candidates. This suggests that there might be a higher possibility for faster rotating TP candidates and B-type stars to evolve into a CBe star.}
    \label{fig:Btype}
\end{figure}

We took a sample of 376 field B-type stars from \cite{2010Huang} and 352 CBe stars from \cite{Yudin2001}. For TP candidates, we considered the sample of 352 synthetically generated sample (obtained through oversampling method, as described in Sect. \ref{sec:ML}). The \textit{v}sin\textit{i} distribution of B-type, CBe stars and TP candidates are shown in Fig. \ref{fig:Btype}. It can be noticed from the figure that B-type stars with high \textit{v}sin\textit{i} values coincide with the distribution of CBe stars and TP candidates. This suggests that there might be a higher possibility for faster rotating TP candidates and B-type stars to evolve into a CBe star. It is also found that the major peak for HAeBe and B-type stars lie in the low \textit{v}sin\textit{i} range by comparing Figs. \ref{fig:Vsini2} and \ref{fig:Btype}. This indicates the possibility of HAeBe stars to become a B-type stars, whereas higher velocity counterparts of HAeBe stars have higher chance of becoming CBe stars. Moreover, we noticed from both the figures that higher \textit{v}sin\textit{i} range is dominated mainly by CBe stars and TP candidates, whereas HAeBe and B-type stars have lesser contribution. Hence, our analysis suggests that the possibility of a TP candidate to become a CBe star is higher than a normal B-type star. Through our analysis, we estimated that 13\% (46 stars) of these TP candidates have a probability of becoming B-type stars, whereas 75\% (264 stars) of the sample have probability to evolve into rapidly rotating CBe stars.

\section{Conclusions}
\label{sec:conclusions}
In the present study, we analyzed a sample of 2167 CBe and 225 HAeBe stars for identifying stars which are in transition from HAeBe to CBe phase. From the optical and infrared photometric analysis we found a sample of 98 stars in transition phase, named as TP candidates. Previous studies reported 14 among these 98 candidates to be HAeBe stars, whereas the rest 84 were classified as CBe stars. To the best of our knowledge, this is the first and also the largest study till date to detect and characterize TP candidates. This identification of TP candidates is done by analyzing the near-IR color-color diagram (which reduced the initial sample to 1297 CBe and 140 HAeBe stars) and by estimating the IR excess emission for these stars in the near-IR regime. From the optical color magnitude diagram we estimated the age and mass of 58 of 98 TP candidates by using MIST isochrones and evolutionary tracks and found the age and mass to be in the range 0.1 -- 5.5 Myr and 2 -- 10.5 $M_\odot$, respectively.

The verification process of the identified TP candidates is performed using the WISE color color diagram, \textit{v}sin\textit{i} distribution study and Machine Learning approach. We found that majority of the identified TP candidates are lying within the range of 0.2 $<$ W1 - W2 $<$ 0.5 and 0.6 $<$ W2 - W3 $<$ 1.1, which is well within the overlapping region of CBe and HAeBe stars. This result suggests that the mid-IR excess observed in the case of TP candidates is similar to that of CBe and HAeBe stars. We found that for HAeBe stars \textit{v}sin\textit{i} ranges within 50 – 125 km s\textsuperscript{-1}, whereas in case of CBe stars \textit{v}sin\textit{i} ranges within 175 – 300 km s\textsuperscript{-1}, respectively. For TP candidates \textit{v}sin\textit{i} show bimodal distribution at 175 -- 225 km s\textsuperscript{-1} and 275 -- 300 km s\textsuperscript{-1}. This implies that \textit{v}sin\textit{i} value is larger for the identified TP candidates than the HAeBe stars and lesser or equal to the upper range of CBe stars.

Furthermore, our oversampling study using machine learning approach also provided satisfactory results. Using the new oversampled dataset, we found that most of the TP candidates are falling in the region between 0.12 $<$ H-Ks $<$ 0.41, which is 93 percent accurate with respect to our previous estimation for TP candidates. We also constructed a new \textit{v}sin\textit{i} distribution plot for 352 CBe, 176 HAeBe and 352 TP candidates as obtained using the new oversampled dataset. From this plot, we observed that although \textit{v}sin\textit{i} for HAeBe stars lies within 50 -- 125 km s\textsuperscript{-1}, but for CBe stars \textit{v}sin\textit{i} ranges within 175 -- 300 km s\textsuperscript{-1}, respectively. Interestingly, for the new synthetically generated TP candidates \textit{v}sin\textit{i} lies majorly around 200 km s\textsuperscript{-1}. This result shows that \textit{v}sin\textit{i} for the TP candidates have shifted slightly towards the intermediate region of HAeBe and CBe stars. 

Lastly, we performed a comparative \textit{v}sin\textit{i} study of the synthetically generated sample of 352 TP candidates with known B-type and CBe stars. Through our analysis, we estimated that 13\% (46 stars) of these TP candidates have a probability of becoming B-type stars, whereas 75\% (264 stars) of the sample have probability to evolve into rapidly rotating CBe stars. Our analysis also suggests that 51 Oph, the star which the provided motivation for this work, is falling under the category of TP candidates. However, as we shift the analysis towards mid-IR color-color diagram, 51 Oph shifts away from the region of TP candidates. Hence, from the analysis we suggest 51 Oph as a more probable TP candidate than a definitive, flag-bearer of the class of TP candidates.

The purpose of this study is to motivate the community of emission-line star research of the significant need of detecting and studying TP candidates in more detail. Due to the fact that the Be phenomenon remains poorly understood after more than 150 years of CBe star research, the identification and characterization of TP candidates may shed new light on the evolutionary phases of various types of emission-line stars. We were unable to conduct spectroscopic investigation on our samples since spectra is currently unavailable for the majority of TP candidates. As a result, we want to perform such a study for our newly identified sample of TP candidates in future utilising high resolution spectra. Spectroscopic investigations of TP candidates, as well as the comparison of their spectral features with that of CBe and HAeBe stars, will aid in the comprehension of their disc properties.

\section*{Acknowledgements}
We would like to thank the Center for research, CHRIST (Deemed to be University), Bangalore, India for providing support. This study has used the Gaia DR2 data to adopt the corresponding distances for program stars. Hence, we express our gratitude to the Gaia collaboration. We also thank the SIMBAD database and the online VizieR library service for helping us in the relevant literature survey. Moreover, I want to thank Mr. Mayang Narang, PhD scholar of Tata Institute for Fundamental Research (TIFR), Mumbai for valuable technical discussions which improved the quality of this work. Furthermore, Mr. Shridharan Bhaskaran and Mr. Arun Roy, both PhD scholars of our department deserve special thanks for their timely suggestions.

%%%%%%%%%%%%%%%%%%%%%%%%%%%%%%%%%%%%%%%%%%%%%%%%%%
\section*{Data Availability}
The data underlying this article are available in 2MASS All Sky Survey, the WISE All Sky Survey, APASS and the CDS VizieR database.
%%%%%%%%%%%%%%%%%%%% REFERENCES %%%%%%%%%%%%%%%%%%

% The best way to enter references is to use BibTeX:

\bibliographystyle{mnras}
\bibliography{example} % if your bibtex file is called example.bib

% Alternatively you could enter them by hand, like this:
% This method is tedious and prone to error if you have lots of references
%\begin{thebibliography}{99}
%\bibitem[\protect\citeauthoryear{Author}{2012}]{Author2012}
%Author A.~N., 2013, Journal of Improbable Astronomy, 1, 1
%\bibitem[\protect\citeauthoryear{Others}{2013}]{Others2013}
%Others S., 2012, Journal of Interesting Stuff, 17, 198
%\end{thebibliography}

%%%%%%%%%%%%%%%%%%%%%%%%%%%%%%%%%%%%%%%%%%%%%%%%%%

%%%%%%%%%%%%%%%%% APPENDICES %%%%%%%%%%%%%%%%%%%%%
\newpage
%\appendix\section{Probable TP candidates}
\begin{landscape}
\centering
\begin{table}
\caption{List of the identified 98 TP candidates from the present study. Spectral type for every star shown in column 4 is adopted from the literature. The corresponding B and V magnitudes are taken from \protect\cite{2015APASS}, whereas the J,H, K$_S$ magnitudes are adopted from \protect\cite{2003Cutri} and W1, W2, W3 magnitudes are obtained from \protect\cite{2014cutri}. The respective A$_V$ values for our stars (presented in column 13) are calculated using the dust extinction maps of \protect\cite{2019Green}. spectral index (n{$_{2-4.6}$}) values for each of these TP candidates is also estimated by us. Previous studies reported 14 (marked with star) among these 98 candidates to be HAeBe stars, whereas the rest 84 were classified as CBe stars.}
\label{tab:table1}
\label{Table1:98 cand}
\begin{tabular}[t]{ccccccccccccccc}
\hline 
\multicolumn{1}{|c|}{\textbf{Simbad ID}}&\multicolumn{1}{c|}{\textbf{RA}}&\multicolumn{1}{c|}{\textbf{Dec}}&\multicolumn{1}{c}{\textbf{Sp.type}}&\multicolumn{1}{|c|}{\textbf{B}}&\multicolumn{1}{c|}{\textbf{V}}&\multicolumn{1}{c|}{\textbf{J}}&\multicolumn{1}{c|}{\textbf{H}}&\multicolumn{1}{c|}{\textbf{K${_s}$}}&\multicolumn{1}{c|}{\textbf{W1}}&\multicolumn{1}{c|}{\textbf{W2}}&\multicolumn{1}{c|}{\textbf{W3}}&\multicolumn{1}{c|}{\textbf{A${_V}$}}&\multicolumn{1}{c|}{\textbf{Spectral index}}&\multicolumn{1}{c|}{\textbf{\textit{v}sin\textit{i}}}\\
\multicolumn{1}{|c|}{\textbf{ }}&\multicolumn{1}{c|}{\textbf{(hh mm ss)}}&\multicolumn{1}{c|}{\textbf{(dd mm ss)}}&\multicolumn{1}{c}{}&\multicolumn{1}{|c|}{\textbf{mag}}&\multicolumn{1}{c|}{\textbf{mag}}&\multicolumn{1}{c|}{\textbf{mag}}&\multicolumn{1}{c|}{\textbf{mag}}&\multicolumn{1}{c|}{\textbf{mag}}&\multicolumn{1}{c|}{\textbf{mag}}&\multicolumn{1}{c|}{\textbf{mag}}&\multicolumn{1}{c|}{\textbf{mag}}&\multicolumn{1}{c|}{\textbf{mag}}&\multicolumn{1}{c|}{\textbf{n{$_{2-4.6}$}}}&\multicolumn{1}{c|}{km s\textsuperscript{-1}}\\
%   & (hh mm ss)  &  (dd mm ss)  &   &  mag & mag &  mag & mag & mag  & mag & mag & mag &  & n{$_{2-4.6}$}  &  km/s \\
\hline
D75b Em* 22-124  	&		22 54 36.62	&		60 48 36.11	&		 B0 	&		16	&		14.4	&		10.4	&		9.8	&		9.4	&		8.7	&		8.3	&		7.1	&		4.1	&		-1.99	&		 --\\		
KAG2008 10573  	&		2 55 2.38	&		60 50 1.71	&		 B8 	&		15.7	&		14.9	&		13	&		12.6	&		12.4	&		11.9	&		11.5	&		9.6	&		1.8	&		-1.91	&		 --\\		
2MASS J00324800+6647596  	&		0 32 48.01	&		66 47 59.55	&		 B3 	&		16.2	&		14.8	&		12.1	&		11.6	&		11.2	&		10.5	&		10.2	&		9.4	&		3.2	&		-1.93	&		 --\\		
2MASS J00412139+6504130  	&		0 41 21.39	&		65 4 13.09	&		 B 	&		14.7	&		13.8	&		12.2	&		11.8	&		11.5	&		10.9	&		10.6	&		9.3	&		2.2	&		-1.97	&		 --\\		
2MASS J00543685+6305498  	&		0 54 36.86	&		63 5 49.84	&		 B2 	&		16	&		15.3	&		13.3	&		13	&		12.6	&		12.1	&		11.8	&		11.1	&		2.2	&		-2.1	&		 --\\		
2MASS J02430559+6316147  	&		2 43 5.60	&		63 16 14.70	&		 B 	&		15.7	&		15	&		13.3	&		13	&		12.7	&		12.2	&		12	&		11.3	&		1.8	&		-2.11	&		 --\\		
2MASS J02455388+6354148  	&		2 45 53.87	&		63 54 14.82	&		 B3 	&		15.6	&		14.6	&		12.5	&		12.1	&		11.8	&		11.3	&		11	&		10.1	&		2.3	&		-2.03	&		 --\\		
2MASS J02475308+6134057  	&		2 47 53.08	&		61 34 5.75	&		 B1 	&		15.8	&		14.8	&		12.6	&		12.2	&		11.9	&		11.4	&		11	&		9.6	&		2.5	&		-1.93	&		 --\\		
2MASS J03013361+5935186  	&		3 1 33.61	&		59 35 18.52	&		 B3 	&		15.5	&		14.6	&		12.4	&		12	&		11.7	&		11.2	&		10.9	&		10	&		2.5	&		-2.1	&		 --\\		
BD-05 4819 	&		18 57 13.03	&		-5 31 30.09	&		 B2 	&		10.7	&		10.3	&		9.2	&		9	&		8.8	&		8.4	&		8.1	&		7.3	&		1.2	&		-2.08	&		 --\\		
BD-10 2073 	&		7 29 58.91	&		-11 15 48.68	&		 B0 	&		10.4	&		10.3	&		9.9	&		9.7	&		9.4	&		9	&		8.7	&		7.7	&		0.2	&		-1.91	&		 --\\		
BD-10 2133 	&		7 37 37.79	&		-10 58 20.41	&		 Be 	&		10.5	&		10.5	&		10.3	&		10.2	&		10	&		9.6	&		9.4	&		8.7	&		0.4	&		-2.13	&		 --\\		
BD-11 2043 	&		7 39 42.24	&		-12 15 33.38	&		 B0 	&		9.8	&		9.7	&		8.9	&		8.8	&		8.5	&		8.1	&		7.7	&		6.8	&		0.3	&		-1.84	&		 --\\		
BD-14 1751 	&		7 10 54.46	&		-14 21 26.22	&		 Be 	&		10.1	&		9.7	&		8.8	&		8.6	&		8.3	&		7.9	&		7.6	&		6.8	&		1	&		-2	&		 --\\		
BD-14 5047 	&		18 26 39.31	&		-14 38 19.36	&		 B0 	&		11	&		10.2	&		8.3	&		8	&		7.6	&		7	&		6.8	&		6	&		1.9	&		-1.95	&		 --\\		
BD-19 1871 	&		7 24 57.48	&		-20 6 58.51	&		 Be 	&		10.5	&		10.1	&		9.2	&		9	&		8.6	&		8.1	&		7.7	&		7	&		1	&		-1.83	&		 --\\		
BD+55 589  	&		2 20 54.94	&		56 23 18.11	&		 B2 	&		10.9	&		10.7	&		10.1	&		9.9	&		9.7	&		9.4	&		9.1	&		8.2	&		0.7	&		-2.18		&	300	{$^{ 5 }$}	\\
BD+56 559  	&		2 21 18.08	&		57 18 22.10	&		 B1 	&		10.4	&		10.1	&		8.6	&		8.4	&		8.1	&		7.7	&		7.5	&		6.7	&		1.2	&		-2.17		&	150	{$^{ 5 }$} \\	
BD+57 607a 	&		2 40 9.23	&		57 45 41.68	&		 Be 	&		11.4	&		10.7	&		9	&		8.7	&		8.2	&		7.7	&		7.2	&		6.3	&		1.6	&		-1.78		&	300	{$^{ 5 }$} 	\\
BD+58 458  	&		2 23 29.55	&		58 57 48.57	&		 Be 	&		10.3	&		9.7	&		8.3	&		8.1	&		7.7	&		7.2	&		6.9	&		6.2	&		1.5	&		-1.91	&		 --\\		
BD+60 2584 	&		23 35 41.50	&		61 11 19.33	&		 O 	&		10.9	&		10.4	&		9.6	&		9.3	&		9	&		8.1	&		7.9	&		7.2	&		2.2	&		-1.72	&		 --\\		
CD-22 4761 	&		7 35 47.51	&		-22 52 38.93	&		 B0 	&		10.5	&		10.1	&		9.1	&		8.9	&		8.7	&		8.1	&		7.8	&		7.1	&		1	&		-1.83	&		 --\\		
CD-23 6121 	&		7 45 8.41	&		-24 7 53.40	&		 Be 	&		10.7	&		10.4	&		9.5	&		9.4	&		9.1	&		8.7	&		8.4	&		7.5	&		1.3	&		-2.09	&		 --\\		
CD-26 4955 	&		7 46 36.20	&		-26 41 40.03	&		 Be 	&		10.7	&		10.4	&		9.5	&		9.3	&		8.9	&		8.4	&		8	&		7.1	&		0.4	&		-1.76	&		 --\\		
CD-27 5181 	&	8 14 3.7	&	-28 19 25.45	&	 Be 	&	11.1	&	10.9	&	10.3	&	10.1	&	9.8	&	9.4	&	9	&	8.1	&	0	&	-1.85	&	 --\\		
CD-28 5235 	&	7 59 22.16	&	-28 54 23.79	&	 B1 	&	10.5	&	10.4	&	9.5	&	9.3	&	9.1	&	8.5	&	8.3	&	7.7	&	0.8	&	-1.93	&	 --\\		
CD-29 6963 	&	8 56 48.25	&	-30 11 1.89	&	 Be 	&	9.6	&	9.3	&	8.6	&	8.4	&	8	&	7.6	&	7.3	&	6.5	&	0.6	&	-1.94	&	 --\\		
CD-30 5559 	&	8 4 28.58	&	-30 58 21.31	&	 Be 	&	10.6	&	10.4	&	9.6	&	9.4	&	9.1	&	8.7	&	8.4	&	7.6	&	0.5	&	-2	&	 --\\		
EM* AS 4 	&	0 44 42.18	&	63 54 27.97	&	 B1 	&	11.3	&	10.6	&	8.8	&	8.5	&	8.2	&	7.8	&	7.5	&	6.9	&	2.3	&	-2.21	&	 --\\		
EM* AS 63  	&	2 33 10.55	&	57 32 54.1	&	 B0 	&	11.7	&	11.1	&	9.6	&	9.3	&	9.1	&	7.6	&	8.4	&	7.8	&	1.6	&	-2.12	&	 --\\		
EM* GGA 150  	&	2 18 2.03	&	56 51 22.02	&	 -- 	&	12.2	&	11.8	&	10.6	&	10.4	&	10.1	&	9.7	&	9.4	&	8.4	&	1	&	-2.02	&	 --\\
EM* GGA 154  	&	2 19 8.63	&	57 3 48.96	&	 B5 	&	12.9	&	12.5	&	11.2	&	10.9	&	10.7	&	10.3	&	10	&	9.1	&	1.2	&	-2.09	&	 --\\
EM* GGA 156  	&	2 19 28.82	&	57 7 4.54	&	 B0 	&	11.6	&	11.1	&	9.6	&	9.5	&	9.2	&	8.8	&	8.6	&	8.3	&	1.2	&	-2.19	&	150	{$^{ 5 }$}	\\
EM* GGA 55 	&	1 33 28.69	&	61 7 59.35	&	 B4 	&	13.3	&	12.8	&	11.8	&	11.5	&	11.3	&	10.6	&	10.3	&	9.5	&	1.2	&	-1.81	&	 --\\		
EM* GGA 56 	&	1 33 41.9	&	60 42 19.67	&	 B2 	&	11.7	&	11.3	&	10.5	&	10.3	&	10.1	&	9.6	&	9.2	&	8.4	&	0.9	&	-1.89	&	333	{$^{ 3 }$} 	\\
EM* GGR 117  	&	22 55 12.34	&	59 7 45.4	&	 B0 	&	14	&	12.9	&	10	&	9.6	&	9.1	&	8.5	&	8.2	&	7.4	&	2.9	&	-1.96	&	 --\\		
EM* GGR 134  	&	23 10 47.51	&	60 31 52.93	&	 B0 	&	13	&	11.8	&	8.9	&	8.4	&	7.9	&	7.3	&	7.1	&	6.4	&	3.2	&	-2.22	&	 --\\

\hline
\end{tabular}
\end{table}
%{\footnotesize\begin{flushleft}
%Full version of this table is made available online.
%\end{flushleft}
%}
\end{landscape}

\begin{landscape}
\centering
\begin{table}
\begin{tabular}[t]{ccccccccccccccc}
\hline \multicolumn{1}{|c|}{\textbf{Simbad ID}}&\multicolumn{1}{c|}{\textbf{RA}}&\multicolumn{1}{c|}{\textbf{Dec}}&\multicolumn{1}{c}{\textbf{Sp.type}}&\multicolumn{1}{|c|}{\textbf{B}}&\multicolumn{1}{c|}{\textbf{V}}&\multicolumn{1}{c|}{\textbf{J}}&\multicolumn{1}{c|}{\textbf{H}}&\multicolumn{1}{c|}{\textbf{K${_s}$}}&\multicolumn{1}{c|}{\textbf{W1}}&\multicolumn{1}{c|}{\textbf{W2}}&\multicolumn{1}{c|}{\textbf{W3}}&\multicolumn{1}{c|}{\textbf{A${_V}$}}&\multicolumn{1}{c|}{\textbf{Spectral index}}&\multicolumn{1}{c|}{\textit{v}sin\textit{i}}\\
\multicolumn{1}{|c|}{\textbf{ }}&\multicolumn{1}{c|}{\textbf{(hh mm ss)}}&\multicolumn{1}{c|}{\textbf{(dd mm ss)}}&\multicolumn{1}{c}{}&\multicolumn{1}{|c|}{\textbf{mag}}&\multicolumn{1}{c|}{\textbf{mag}}&\multicolumn{1}{c|}{\textbf{mag}}&\multicolumn{1}{c|}{\textbf{mag}}&\multicolumn{1}{c|}{\textbf{mag}}&\multicolumn{1}{c|}{\textbf{mag}}&\multicolumn{1}{c|}{\textbf{mag}}&\multicolumn{1}{c|}{\textbf{mag}}&\multicolumn{1}{c|}{\textbf{mag}}&\multicolumn{1}{c|}{\textbf{n{$_{2-4.6}$}}}&\multicolumn{1}{c|}{km s\textsuperscript{-1}}\\
 %& (hh mm ss)  &  (dd mm ss)  &   &  mag & mag &  mag & mag & mag  & mag & mag & mag &  & n{$_{2-4.6}$}  &  km/s \\
\hline
EM* GGR 190  	&	0 32 44.56	&	63 18 16.75	&	 B8 	&	12.3	&	11.9	&	10.5	&	10.1	&	9.7	&	9.2	&	8.9	&	8.3	&	1.7	&	-2.07	&	 --\\		
EM* GGR 50 	&	22 12 19.64	&	57 16 5.38	&	 B0 	&	13	&	12.2	&	10.3	&	10	&	9.7	&	9.1	&	8.8	&	8	&	2.4	&	-2.05	&	 --\\		
EM* GGR 63 	&	22 20 10.11	&	58 6 34.37	&	 B1 	&	14.1	&	13.5	&	11.2	&	10.8	&	10.5	&	10.1	&	9.9	&	9.1	&	1.4	&	-2.16	&	 --\\		
EM* MWC 1015 	&	20 21 33.58	&	37 24 51.7	&	 B1 	&	13.4	&	11.9	&	8.5	&	8	&	7.6	&	6.9	&	6.7	&	5.2	&	4	&	-2.12	&	 --\\		
EM* MWC 447  	&	2 22 46.97	&	56 58 5.89	&	 B0 	&	11	&	10.6	&	9.5	&	9.2	&	8.9	&	8.6	&	8.3	&	7.3	&	1.1	&	-2.16	&	300 {$^{ 5 }$}	\\	
EM* MWC 465  	&	3 53 58.25	&	46 53 51.77	&	 B0 	&	10.7	&	10.3	&	9.3	&	9	&	8.8	&	8.3	&	8	&	7.1	&	1.1	&	-1.93	&	 --\\		
EM* MWC 659  	&	22 47 45.37	&	57 16 50.8	&	 B0 	&	10.7	&	10.2	&	8.8	&	8.6	&	8.2	&	7.8	&	7.4	&	6.5	&	1.2	&	-1.96	&	 --\\		
EM* MWC 674  	&	0 42 26.48	&	64 3 0.17	&	 B1 	&	11.5	&	10.6	&	8.5	&	8.1	&	7.8	&	7.3	&	7	&	6.3	&	2.4	&	-2.1	&	 --\\		
EM* MWC 678  	&	0 49 48.2	&	60 59 10.01	&	 B2 	&	10.6	&	10	&	8.8	&	8.7	&	8.4	&	7.8	&	7.4	&	6.6	&	1.3	&	-1.76	&	 --\\		
EM* RJHA 40  	&	6 42 29.78	&	0 53 58.21	&	 B3 	&	14	&	13.2	&	11	&	10.6	&	10.3	&	9.7	&	9.3	&	8.4	&	2.2	&	-1.92	&	 --\\		
EM* UHA 141  	&	21 11 0.95	&	45 39 41.33	&	 B1 	&	14.1	&	13.7	&	12.1	&	11.9	&	11.6	&	11.3	&	11	&	10.2	&	1.6	&	-2.19	&	 --\\		
HBHA 6210-03 	&	1 9 58.8	&	62 52 29.25	&	 B3 	&	15.6	&	14.5	&	12.4	&	12	&	11.6	&	10.8	&	10.6	&	10	&	2.9	&	-1.88	&	 --\\		
HD 15450 	&	2 31 19.53	&	56 53 52.18	&	 B1 	&	9	&	8.7	&	8.2	&	8	&	7.7	&	7.1	&	6.8	&	5.9	&	1	&	-1.79	&	242 {$^{ 5 }$}	\\	
HD 160886  	&	17 43 11.21	&	-18 18 8.37	&	 B3 	&	10	&	9.8	&	9	&	8.9	&	8.7	&	8.3	&	8.1	&	7.4	&	0	&	-2.04	&	328	{$^{ 6 }$} 	\\
HD 161103  	&	17 44 45.77	&	-27 13 44.48	&	 B0 	&	8.8	&	8.4	&	7.1	&	6.8	&	6.5	&	5.8	&	5.6	&	5	&	1.4	&	-1.88	&	260	{$^{ 6 }$} 	\\
HD 161306  	&	17 45 7.7	&	-9 48 54.45	&	 B3 	&	9	&	8.4	&	6.9	&	6.6	&	6.2	&	6	&	5.6	&	4.8	&	1.7	&	-2.2	&	 --\\		
HD 169805  	&	18 27 12.01	&	-18 57 13.4	&	 B2 	&	8.2	&	7.9	&	7.4	&	7.3	&	7	&	6.4	&	6.3	&	5.5	&	0.6	&	-1.96	&	274 {$^{ 6 }$}	\\	
HD 170061  	&	18 28 0.74	&	-14 42 24.62	&	 OB 	&	9.8	&	9.4	&	8.1	&	7.8	&	7.4	&	6.9	&	6.6	&	6	&	0.9	&	-1.89	&	 --\\		
HD 172694  	&	18 42 16.57	&	-15 51 20.83	&	 B2 	&	8.4	&	8.1	&	7.4	&	7.2	&	6.9	&	6.4	&	6	&	5.2	&	0.8	&	-1.81	&	 --\\
HD 184279  	&	19 33 36.92	&	3 45 40.79	&	 B0 	&	7.7	&	7.3	&	6.9	&	6.7	&	6.4	&	6	&	5.6	&	4.9	&	0.2	&	-1.74	&	212	{$^{ 6 }$}	\\
HD 187350  	&	19 49 33.45	&	-1 6 3.72	&	 B1 	&	8.5	&	8.4	&	7.7	&	7.7	&	7.4	&	7.1	&	6.8	&	6	&	0.5	&	-2.09	&	199	{$^{ 6 }$}	\\
HD 205618  	&	21 35 44.48	&	29 44 43.89	&	 B2 	&	8.6	&	8.1	&	7.9	&	7.8	&	7.5	&	7	&	6.7	&	5.8	&	0.3	&	-1.93	&	299	{$^{ 6 }$}	\\
HD 227611  	&	20 5 45.15	&	35 54 2.99	&	 B1 	&	9.2	&	8.8	&	7.7	&	7.5	&	7.2	&	6.6	&	6.2	&	5.4	&	1.2	&	-1.78	&	 --\\		
HD 227836  	&	20 8 4.56	&	36 7 26.04	&	 B5 	&	10.8	&	10.5	&	9.2	&	8.9	&	8.5	&	8.3	&	7.9	&	6.9	&	1.1	&	-2.21	&	 --\\		
HD 229221  	&	20 23 45.96	&	38 30 3.15	&	 B0 	&	10.2	&	9.2	&	7.1	&	6.7	&	6.3	&	6.1	&	5.6	&	5.1	&	2.2	&	-2.17	&	 --\\		
HD 254647  	&	6 17 34.62	&	11 11 27.31	&	 Bp 	&	10.4	&	10.1	&	8.9	&	8.6	&	8.3	&	7.9	&	7.6	&	6.9	&	1.3	&	-2.03	&	 --\\		
HD 277707  	&	5 12 11.62	&	39 33 20.38	&	 Bp 	&	10.8	&	10.3	&	9.1	&	8.9	&	8.7	&	8.3	&	8.1	&	7.3	&	1.1	&	-2.15	&	 --\\		
HD 35347 	&	5 25 17.82	&	29 36 53.6	&	 B1 	&	9.2	&	8.9	&	8.1	&	7.9	&	7.8	&	7.5	&	7.2	&	6.4	&	0.9	&	-2.2	&	 --\\		
HD 39340 	&	5 53 6.09	&	26 26 43.44	&	 B3 	&	8.3	&	8.1	&	7.6	&	7.6	&	7.3	&	7	&	6.7	&	5.9	&	0.7	&	-2.2	&	222	{$^{ 6 }$}	\\
HD 46380 	&	6 32 43.23	&	-7 30 32.28	&	 B2 	&	8.4	&	8.1	&	6.9	&	6.7	&	6.4	&	5.9	&	5.5	&	4.8	&	1.2	&	-1.77	&	290	{$^{ 5 }$}	\\
HD 49977 	&	6 50 26.47	&	-14 6 48.12	&	 B5 	&	8.2	&	8	&	7.4	&	7.2	&	6.9	&	6.6	&	6.2	&	5.5	&	0.7	&	-2.01	&	218	{$^{ 5 }$}	\\
HD 50083 	&	6 51 45.75	&	5 5 3.86	&	 B3 	&	7	&	6.9	&	6.6	&	6.5	&	6.3	&	6	&	5.4	&	4.7	&	0.2	&	-1.78	&	176	{$^{ 4 }$}	\\
HD 57775 	&	7 22 24.83	&	-10 49 21.48	&	 Be 	&	9.1	&	8.8	&	8.3	&	8.1	&	7.8	&	7.4	&	7.1	&	6.5	&	0.9	&	-2.08	&	 --\\		
IPHAS J004353.19+595641.6  	&	0 43 53.18	&	59 56 41.64	&	 B1 	&	16.1	&	15.6	&	14	&	13.8	&	13.4	&	13.1	&	12.8	&	11.3	&	1.3	&	-2.21	&	 --\\		
IPHAS J011543.99+660116.2  	&	1 15 43.93	&	66 1 16.18	&	 B3 	&	16.2	&	14.9	&	12	&	11.5	&	11.1	&	10.5	&	10.1	&	9.1	&	2.5	&	-1.89	&	 --\\		
IPHAS J012320.11+635830.9  	&	1 23 20.1	&	63 58 30.68	&	 B3 	&	16	&	14.7	&	11.9	&	11.4	&	11	&	10.5	&	10.1	&	9.2	&	3	&	-2.06	&	 --\\		
IPHAS J020325.84+584145.0  	&	2 3 25.84	&	58 41 44.82	&	 B3 	&	15.7	&	15.2	&	14.5	&	14.2	&	14	&	13.1	&	12.9	&	11.2	&	1.3	&	-1.66	&	 --\\		
IPHAS J022107.83+625754.6  	&	2 21 7.82	&	62 57 54.55	&	 B 	&	16.2	&	15.4	&	13.2	&	12.9	&	12.6	&	12.1	&	11.8	&	10.8	&	2.2	&	-2.15	&	 --\\		
IPHAS J024252.56+611953.9  	&	2 42 52.58	&	61 19 53.87	&	 B3 	&	17.1	&	16.2	&	14	&	13.6	&	13.2	&	12.9	&	12.5	&	11.9	&	2.4	&	-2.22	&	 --\\		
MR 8 	&	7 43 13.07	&	-29 18 33.79	&	 OB 	&	12.8	&	12.4	&	11.2	&	10.9	&	10.6	&	10.2	&	9.9	&	9	&	1.3	&	-2.03	&	 --\\		
NGC 1893 35  	&	5 22 43.01	&	33 25 5.42	&	 B1 	&	14.4	&	13.6	&	11.6	&	11.3	&	10.9	&	10.4	&	9.9	&	8.1	&	2.2	&	-1.82	&	 --\\

\hline
\end{tabular}
\end{table}
\end{landscape}

\begin{landscape}
\centering
\begin{table}
\begin{tabular}[t]{ccccccccccccccc}
\hline \multicolumn{1}{|c|}{\textbf{Simbad ID}}&\multicolumn{1}{c|}{\textbf{RA}}&\multicolumn{1}{c|}{\textbf{Dec}}&\multicolumn{1}{c}{\textbf{Sp.type}}&\multicolumn{1}{|c|}{\textbf{B}}&\multicolumn{1}{c|}{\textbf{V}}&\multicolumn{1}{c|}{\textbf{J}}&\multicolumn{1}{c|}{\textbf{H}}&\multicolumn{1}{c|}{\textbf{K${_s}$}}&\multicolumn{1}{c|}{\textbf{W1}}&\multicolumn{1}{c|}{\textbf{W2}}&\multicolumn{1}{c|}{\textbf{W3}}&\multicolumn{1}{c|}{\textbf{A${_V}$}}&\multicolumn{1}{c|}{\textbf{Spectral index}}&\multicolumn{1}{c|}{\textit{v}sin\textit{i}}\\
\multicolumn{1}{|c|}{\textbf{ }}&\multicolumn{1}{c|}{\textbf{(hh mm ss)}}&\multicolumn{1}{c|}{\textbf{(dd mm ss)}}&\multicolumn{1}{c}{}&\multicolumn{1}{|c|}{\textbf{mag}}&\multicolumn{1}{c|}{\textbf{mag}}&\multicolumn{1}{c|}{\textbf{mag}}&\multicolumn{1}{c|}{\textbf{mag}}&\multicolumn{1}{c|}{\textbf{mag}}&\multicolumn{1}{c|}{\textbf{mag}}&\multicolumn{1}{c|}{\textbf{mag}}&\multicolumn{1}{c|}{\textbf{mag}}&\multicolumn{1}{c|}{\textbf{mag}}&\multicolumn{1}{c|}{\textbf{n{$_{2-4.6}$}}}&\multicolumn{1}{c|}{km s\textsuperscript{-1}}\\
% & (hh mm ss)  &  (dd mm ss)  &   &  mag & mag &  mag & mag & mag  & mag & mag & mag &  & n{$_{2-4.6}$}  &  km/s \\
\hline
		
NGC 2345 24  	&	7 8 11.62	&	-13 9 26.94	&	 B3 	&	14.9	&	13.9	&	11.6	&	11.3	&	11	&	10.5	&	10.4	&	9.8	&	1.8	&	-2.16	&	 --\\
NGC 6649 9 	&	18 33 28.3	&	-10 24 8.76	&	 B0 	&	13	&	11.8	&	8.7	&	8.2	&	7.8	&	7.3	&	7	&	6.1	&	2.5	&	-2.05	&	 --\\		
Schulte 64 	&	20 33 18.53	&	41 15 35.15	&	 B1 	&	17.6	&	15	&	9.4	&	8.6	&	8	&	7.2	&	6.8	&	6.2	&	4.8	&	-1.91	&	 --\\		
UCAC3 274-184438 	&	20 35 41.52	&	46 46 49.53	&	 B0 	&	13.8	&	12.8	&	10.2	&	9.7	&	9.3	&	8.9	&	8.5	&	7	&	2.4	&	-2.03	&	 --\\		
UCAC3 297-175444 	&	22 19 51.45	&	58 8 53.47	&	 B0 	&	13.8	&	12.7	&	10.1	&	9.7	&	9.4	&	8.9	&	8.6	&	7.7	&	3	&	-2.19	&	 --\\		
UCAC4 448-022358 	&	6 45 3.44	&	0 34 13.92	&	 K3 	&	14.8	&	14	&	12	&	11.7	&	11.4	&	11	&	10.7	&	10	&	1.5	&	-2.21	&	 --\\		
AS 477 *	&	21 52 34.1	&	47 13 43.6	&	 A0 	&	10.4	&	10.1	&	8.5	&	8	&	7.2	&	5.4	&	4.6	&	3.5	&	4.7	&	-0.07	&	150	{$^{ 5 }$}	\\
GSC 6546-3156 *	&	7 24 17.55	&	-26 16 5.3	&	 A0 	&	14.8	&	14.1	&	12.3	&	11.9	&	11.6	&	10.3	&	9.8	&	4.7	&	2.5	&	-0.8	&	 --\\		
HBC 217 *	&	6 40 42.18	&	9 33 37.42	&	 G0 	&	12.5	&	12	&	10.8	&	10.3	&	9.8	&	9	&	8.5	&	6.5	&	3.8	&	-1.57	&	 --\\		
HBC 222 *	&	6 40 51.19	&	9 44 46.11	&	 F8 	&	12.6	&	12	&	10.7	&	10.2	&	9.7	&	8.7	&	8.3	&	7.1	&	5.1	&	-1.75	&	 --\\		
HD 290500 *	&	5 29 48.05	&	0 23 43.51	&	 A2 	&	11.3	&	11	&	10.3	&	10	&	9.6	&	8.3	&	7.8	&	5.6	&	0.8	&	-0.74	&	84 {$^{ 1 }$}	\\	
HD 35929 *	&	5 27 42.79	&	-8 19 38.45	&	 F2 	&	8.5	&	8.1	&	7.2	&	7	&	6.7	&	6.2	&	5.5	&	4.2	&	0.4	&	-1.37	&	 --\\		
EM* MWC 1021 *	&	20 29 26.92	&	41 40 43.9	&	 Bp 	&	15.6	&	12.8	&	6.5	&	5.5	&	4.6	&	2.4	&	0.5	&	0.4	&	8.6	&	1.24	&	 --\\		
PDS 22 *	&	6 3 37.07	&	-14 53 3.17	&	 A0 	&	10.3	&	10.2	&	9.9	&	9.7	&	9.2	&	8.1	&	7.3	&	3.3	&	0.9	&	-0.54	&	 --\\		
PDS 124 *	&	6 6 58.5	&	-5 55 6.7	&	 A0 	&	12.9	&	12.5	&	11.1	&	10.6	&	9.8	&	8.4	&	7.7	&	4.3	&	3.8	&	-0.6	&	 --\\		
PDS 129 *	&	6 31 3.64	&	10 1 13.51	&	 F5 	&	12.8	&	12.1	&	10.5	&	10.1	&	9.7	&	9	&	8.4	&	5.3	&	1.8	&	-1.4	&	 --\\		
PDS 469 *	&	17 50 58.1	&	-14 16 11.86	&	 A0 	&	13.4	&	12.8	&	11.2	&	10.8	&	10.4	&	9.6	&	9.2	&	4.9	&	1.9	&	-1.58	&	 --\\		
V1787 Ori *	&	5 38 9.3	&	-6 49 16.6	&	 A5 	&	15.3	&	13.8	&	9.9	&	9	&	8	&	6.5	&	5.7	&	3.1	&	9.1	&	-1	&	390	{$^{ 1 }$} \\	
V380 Ori *	&	5 36 25.43	&	-6 42 57.68	&	 A1 	&	11.3	&	10.6	&	8.1	&	7	&	5.9	&	4.8	&	3.7	&	1.5	&	9.5	&	-1.15	&	 --\\
V599 Ori *	&	5 38 58.64	&	-7 16 45.63	&	 A8 	&	15.4	&	13.7	&	9.6	&	8.6	&	7.6	&	6.7	&	6	&	4.3	&	8	&	-1.77	&	66 {$^{ 2 }$}	\\

\hline
\end{tabular}
{\footnotesize\begin{flushleft}
{${^1}$\cite{2019Kounkel};
${^2}$\cite{2016DaRio}; 
${^3}$\cite{2010Huang}; 
${^4}$\cite{2005fremat}; 
${^5}$\cite{2005Glecbocki};
${^6}$\cite{Yudin2001}}
\end{flushleft}
}
%\tablefootnote{${^1}$\cite{2019Kounkel};
%${^2}$\cite{2016DaRio}; 
%${^3}$\cite{2010Huang}; 
%${^4}$\cite{2005fremat}; 
%${^5}$\cite{2005Glecbocki};
%${^6}$\cite{Yudin2001}}
\end{table}
\bsp	% typesetting comment
\label{lastpage}
\end{landscape}
\end{document}